\documentclass[prb,twocolumn,showpacs,superscriptaddress,amsmath,amssymb]{revtex4}

\usepackage{graphicx,color}

\usepackage{dcolumn}
\usepackage{bm}
\begin{document}
%
\title{Comprehensive study of sodium, copper, and silver clusters\\
over a wide range of sizes $2\leq N\leq 75$}
%
%
%
%
%
%
%
%
\author{Masahiro Itoh}
\email{itoh.japan@gmail.com}
\affiliation{Institute for Materials Research, Tohoku University, Aoba-ku, Sendai 980-8577, Japan}
\affiliation{Institute of Multidisciplinary Research for Advanced Materials, Tohoku University, Aoba-ku, Sendai 980-8577, Japan}

\author{Vijay Kumar}
\affiliation{Institute for Materials Research, Tohoku University, Aoba-ku, Sendai 980-8577, Japan}
\affiliation{Dr.Vijay Kumar Foundation, 45 Bazaar Street, K. K. Nagar (West), Chennai 600 078, India}
%
%
%
\author{Tadafumi Adschiri}
\affiliation{Institute of Multidisciplinary Research for Advanced Materials,
Tohoku University, Aoba-ku, Sendai 980-8577, Japan}
\affiliation{Advanced Institute for Materials Research, WPI, Tohoku University, Aoba-ku, Sendai 980-8577, Japan}
\author{Yoshiyuki Kawazoe}
\affiliation{Institute for Materials Research, Tohoku University, Aoba-ku, Sendai 980-8577, Japan}
\date{\today}

\begin{abstract}
The geometric and electronic structures of Na$_N$, Cu$_N$, and Ag$_N$ metal clusters 
are systematically studied based on the density functional theory over a wide range of cluster sizes $2\leq N\leq 75$.
A remarkable similarity is observed between the optimized geometric structures
of alkali and noble metal clusters over all of the calculated cluster sizes $N$.
The most stable structures are the same for the three different metal clusters for approximately half the cluster sizes $N$ considered in this study.
Even if the most stable structures are different, the same types of structures are obtained when the meta-stable structures are also considered.
For all of the three different metal clusters, the cluster shapes change in the order of linear, planar, opened, and closed structures with increasing $N$.
This structural type transition leads to a deviation from the monotonic increase in the volume with $N$.
A remarkable similarity is also observed for the $N$ dependence of the cluster energy $E(N)$ for the most stable geometric structures.
The amplitude of this energy difference is larger in the two noble metal clusters than in the alkali metal cluster.
This is attributed to the contribution of $d$ electrons to the bonds.
The magic number is defined in the framework of total energy calculations for the first time.
In the case of Na$_N$, a semi-quantitative comparison between the experimental abundance spectra
(Knight \textit{et al}., Phys. Rev. Lett. \textbf{52}, 2141 (1984)) and the total energy calculations is carried out.
The changing aspect of the Kohn-Sham eigenvalues from $N=2$ to $N=75$ is presented for the three different metal clusters.
The feature of the bulk density of states already appears at $N=75$ for all of three clusters.
With increasing $N$, the HOMO-LUMO gap clearly exhibits an odd-even alternation and converges to 0.
Although there is a similarity in the $N$ dependence of the HOMO-LUMO gap between the three metal clusters,
it is much stronger between the two noble metal clusters.
The growth aspect of the $d$ band below the Fermi level of the noble metal clusters with increasing $N$ is presented.
A good correspondence is observed in the $d$ characteristic of the electronic states between the cluster composed of 75 atoms and the bulk metal.
The similarities observed in the $N$ dependence of the geometric structures and $E(N)$s originate from the similarity in that of the electronic structures. 
\end{abstract}

\maketitle
\section{Introduction}
It is beneficial to study the geometric structures and various electronic properties of clusters
as an intermediate phase of materials between isolated and condensed systems;
the combination of quantum chemistry and solid state physics will result in further developments in the field of materials science.
In the basis of the density functional theory (DFT) \cite{HK, KS}, an \textit{ab initio} many-body theory for systems in the ground state,
a detailed comparative study of the most and meta-stable geometric and electronic structures
of three different metal clusters - Na$_N$, Cu$_N$, and Ag$_N$ - is systematically carried out over a wide range of $2\leq N\leq 75$.
The three types of atoms composing those clusters
have a common feature in that the outermost valence electron is one $s$ electron.
The changing and converging aspects of the geometric and electronic structures
of alkali and noble metal clusters from a diatomic molecule (dimer)
to a condensed system have been presented explicitly and compared for the first time.
A simple review of previous studies on clusters is presented below.

In 1984, a study was carried out based on the Woods-Saxon type shell model \cite{Knight}
to explain the distinctive peak observed in the abundance spectra of a Na$_N$ cluster
at special value of $N$ ($N=8, 20, 40, 58, 92$, ...) evaluated in an experiment.
Using this shell model, the electronic structures of simple metal clusters based on
a simple spherical effective potential using phenomenological parameters.
In this model, the Coulomb interactions between electrons are completely neglected.
In a later study, the anormalous $N$-dependent stability of a Na$_N$ cluster was studied
based on various types of shell models such as a harmonic oscillator, with the effect of Coulomb interactions being neglected \cite{Clemenger}.
%
%

In addition to simple shell models, various types of jellium models \cite{Chou, Ekardt-1984, UJM} were adopted for Na$_N$ clusters.
In jellium models for metal clusters, the equilibrium ionic configuration of a cluster is averaged
and replaced by a uniform or nearly uniform potential with a surface boundary.
In most previous studies, the jellium models were incorporated with the DFT and the total energies were evaluated.

In model-based studies of Na$_N$, the $N$-dependent cluster stability was evaluated based on the shell correction energy,
curvature of the sum of the electronic eigenvalues,
and total energy \cite{Brack-RMP-1, LDM, Nishioka, Bjornholm, Genzken, Brack-RMP-2, deHeer-RMP, Chou, Ekardt-1984, UJM, Kumar-Kawazoe-review}.
For the evaluation of the $N$-dependent total energy, a semi-quantitative comparison
with the peak intensity $I(N)$ in the experimental abundance spectra \cite{Knight-deHeer} was also carried out.

However, cluster studies based on these models differ from those based on the first principles calculations,
which directly relate the geometric structures to the electronic structures.
Even if the experimental magic number for a particular value of $N$ is obtained accidentally,
these models are not based on a close relationship between the geometric and electronic structurtes of real Na$_N$ clusters.
Furthermore, using these models, it is impossible to study the similarities and differences
between the alkali metal cluster Na$_N$ and the noble metal clusters Cu$_N$ and Ag$_N$
because the $s$ electron in the outermost shell as well as $d$ electrons in the inner shell contribute to bonding in the latter.

Recenly, the most and meta-stable structures of Na$_N$, Cu$_N$, and Ag$_N$ clusters
were determined for $N\leq 22$ from first principles calculations based on DFT \cite{3-Kronig, Itoh-1, ref-Itoh-1, Yang-1, Yang-2}.
In this study, an extremely systematic comparative study based on DFT was carried out for the three different metal clusters up to $N=75$.
Our study successfully reproduced the results of stable geometric structures obtained in previous studies;
furthermore, it successfully discovered the convergence aspects of the electronic structures of clusters with those of bulk metals.

In section II, the computational method and the approximation are described.
In section III, for the most and meta-stable structures are systematically classified
and the similarities and differences between the three different metal clusters are described.
In section IV, the $N$ dependence of the ground state energy of the most stable cluster structure is analyzed.
In section V, all aspects of the electronic structure of the clusters including the odd-even alternation of the HOMO-LUMO gap are described.
In section VI, the conclusion of this study is presented.
\section{Computational Methods}
%
%
An exact many-body theory for a ground state system DFT \cite{HK, KS} -
was employed to study the geometric and electronic structures of Na$_N$, Cu$_N$, and Ag$_N$ clusters.
In this theory, the many-body Schr\"{o}dinger equation is transformed
into a self-consistent Schr\"{o}dinger - type equation for a virtual one - electron in the reference system.
This equation is called the Kohn-Sham equation and it is derived from the Hohenberg-Kohn variational principles,
namely, $\frac{\delta E[\rho(\textbf{r})]}{\delta \rho(\textbf{r})}=0$.
In the DFT, various physical properties of materials such as the exact ground state total energy can be evaluated
if the exact exchange-correlation energy functional is adopted.
In principle, the exact ionization energy for a system can be evaluated as the highest occupied molecular orbital energy by solving the Kohn-Sham equation.
The other orbital energies are considered to approximately describe the quasiparticle energy.

In this study, the generalized gradient approximation (GGA) proposed by Perdew and Wang (PW91) \cite{PW91} is employed
for the exchange-correlation energy functional. The spin polarization of the system was also considered.
In the local density approximation (LDA), which was the first approximation developed for the exchange-correlation energy functional,
the energy functional for a real inhomogeneous system is locally approximated by an energy functional of a homogeneous electron liquid.
The main disadvantage of the LDA is the incomplete cancellation of the self-interaction fused into the Hartree potential by the exchange potential.
This leads to an incorrect description of the electronic properties when electrons are strongly localized in the system.
In fact, the GGA was developed to overcome this disadvantage of the LDA.
It improves upon the approximation of several physical properties such as interatomic distance and cohesive energy in many systems.

In order to reduce the number of plane waves used in the calculation of the electronic structure,
a pseudopotential method is employed for an approximation of the nucleus, inner core electrons, and valence electrons.
An unempirical pseudopotential called ultrasoft pseudopotential, which is known to have good transferability \cite{Vanderbilt}, was employed.
In a manner similar to other methods, relativistic correction terms such as the mass velocity and the Darwin terms are included in the pseudopotentials.
For Na, Cu, and Ag atoms, 3$s^{1}$, $3d^{10}4s^{1}$, and $4d^{10}5s^{1} $electrons were explicitly treated as the valence electrons, respectively.

A local orbital basis set is commonly adopted within a real space representation
for the electrons of isolated systems such as atoms, molecules, and clusters.
However, it is difficult to converge the total energy effectively
because some expertise is required to select the basis function for each atom.

Instead of treating the isolated system as is, the total energy can be effectively converged
by using a plane wave basis set within a reciprocal space representation, which is frequently used in periodic systems.
Here, isolated systems are approximated by pseudo-crystal systems constructed using arranged isolated systems having sufficient space between them.

In this method, the total energy can be effectively converged by simply increasing the cut-off energy of a plane wave expansion.
The total energy of the cluster is approximately evaluated from the total energy of the pseudo-crystal system per unit cell.
Another advantage of employing this method is that
it is possible to evaluate the total energies and electronic structures of the clusters and bulk solids using a single method.

However,
the latter method has two disadvantages in the study of isolated systems. 
The first is the incompleteness of removing the interaction between isolated systems.
The second is the inability to separate the total energy into
the kinetic energy term of electrons, and potential terms between electron and electron, electron and nucleus, and nucleus and nucleus.
This is because the Ewald method \cite{Ewald} is adopted for the evaluation of the total energy.
In this method, the total energy is effectively converged by separating the Coulomb potential into the short- and long-range parts,
and then treating the former in the real space and the latter in the reciprocal space.

The edge length of each cubic unit cell of the pseudo-crystal systems was set
to be 20$\sim$30 \AA\ for Na$_N$ clusters, and 15$\sim$30 \AA\ for Cu$_N$ and Ag$_N$ clusters.
It was confirmed that these lengths are sufficiently large to neglect the electronic wave vector dependence in these electronic structures.
The cut-off energy for the plane wave expansion was set to be 48.7, 233.7, and 180.7 eV for Na$_N$, Cu$_N$, and Ag$_N$, respectively.
%
%
Brillouin zone integration was carried out only for the $\Gamma$-point in the pseudo-crystal system.
This \textit{\textbf{k}}-point selection is known to be the most effective in the calculation of an isolated system.
The residual minimization scheme, direct inversion in the iterative subspace algorithm
was adopted for the effective self-consistent calculation of the electronic structures \cite{RMM-DIIS-1, RMM-DIIS-2}.
The convergence criterion of the total energy was set to be within 1 $\times$ $10^{-4}$ eV.

Geometric structures of the clusters were optimized from several hypothetical initial structures
using an optimization algorithm called the conjugate gradient method \cite{Payne}.
As initial structures in the geometry optimization procedure for Na$_N$ and Cu$_N$,
the optimized cluster structures obtained from empirical atomic pair potentials such as Lennard-Jones, Morse, Sutton-Chen, etc. \cite{Cambridge},
were employed along with optimized structures based on first principles calculations that were obtained in previous studies \cite{Itoh-1, ref-Itoh-1, Yang-1}.

In addition to these structures, structures expected to exist
from the experimental photoelectron spectra \cite{Wrigge, Issendorff-private, Hakkinen-2004, Kostko-2005, Kostko-2007}
and those expected to exist in the neighborhood of the local minimum points on the potential energy surface
from our experience of the optimization procedures for these systems \cite{Itoh-1} were selected.
The optimized structures of Na$_N$ and Cu$_N$ obtained in this study were employed
as the initial structures of Ag$_N$ in the geometry optimization procedure,
and the interatomic distances satisfying the nearest neighbor atomic distance ratio between the Na, Cu, and Ag bulk crystals were scaled.

The electronic density of states of the Na, Cu, and Ag bulk crystals were calculated for comparison with those of the clusters.
The same cut-off energies and optimization method used for the cluster systems were also used for the bulk calculations.
The lattice constants were optimized for the hcp and bcc structures for Na, and the fcc structure for Cu and Ag.
Brillouin zone integration was carried out for the \textit{\textbf{k}}-point meshes generated by Monkhorst-Pack scheme \cite{Monkhorst}.
For the bulk bcc, fcc, and hcp structures,
the number of meshes were selected to be 8 $\times$ 8 $\times$ 8, 8 $\times$ 8 $\times$ 8, and 8 $\times$ 8 $\times$ 4, respectively.
The convergence criteria of the total energy for the bulk bcc, fcc, and hcp structure were set to be
within 1.0 $\times$ $10^{-4}$, 1.0 $\times$ $10^{-4}$, and 5.0 $\times$ $10^{-3}$ eV/atom, respectively.

The nearest neighbor diatomic distances, binding energies, and bulk moduli for the dimers and bulk crystals are presented, in the Appendix.

The Vienna Ab-initio Simulation Package (VASP) was employed in this study \cite{Kresse}.

\section{Cluster Geometries}
\subsection{$N$ dependence of most and meta-stable structures}
%
%
In general, unlike the case of molecules, clusters have various energetically nearly degenerated meta-stable structures.
Therefore, the meta-stable as well as most stable structures must be considered for understanding cluster-related phenomena.

Figures 1, 2, and 3 show the most and meta-stable structures of the Na$_N$, Cu$_N$, and Ag$_N$ clusters
for $3 \leq N\leq 12$, $13 \leq N\leq 22$, and $34 \leq N\leq 75$, respectively.
In these figures, the meta-stable structures are carefully selected to show the structural type transition, as described below.
Almost all of the most and meta-stable structures of the Na$_N$, Cu$_N$, and Ag$_N$ clusters are similar.
Therefore, to save space, the cluster structures are represented by the structure of Cu$_N$ in the figure.
The types of these structures are classified by the notations I, II, III, IV, and V.
This classification obeys the order of stability in the structural type of Cu$_N$.
The symbols L, P, O, and C after I, II, III, IV, and V indicate linear, planar, opened, and closed structures, respectively.
For a detailed analysis of the correlation between the structural type and the energetical stability of the clusters,
these structural types should be further classified according to the symbols.
Both opened and closed structures are three-dimensional.
The coordination number (CN) is a necessary concept for the classification of these three-dimensional structures.
Although the CN can also be defined for linear and planar structures, it is not required for the classification in this study.
An opened structure is defined as one without atoms whose CNs are greater than or equal to 11.
Other three-dimensional structures are defined as closed structures.
The three values listed after the symbols L, P, O, and C represent the relative energies of
the most stable structures of Na$_N$, Cu$_N$, and Ag$_N$, respectively.
A value of 0.000 is assigned to the most stable structure,
and the relative energies of the meta-stable structures are expressed in electron volts.
We consider the case of $N=6$ as an example.
As shown in FIG.1, the structural type of the most stable structure in Na$_6$ is II, while that in Cu$_6$ and Ag$_6$ is I.
The structural types of the 2nd stable structures I in Na$_6$ are I and in Cu$_6$ and Ag$_6$, II.
The structural type of the 3rd stable structures in Cu$_6$ and Ag$_6$ is III.
However, structure III is not found in Na$_6$. A symbol N indicates a structural type that has not been found.
The two different figures of structure II show the same cluster structure viewed from different angles.
Such different figures for the same structure are also shown for the other values of $N$.
Structural type IV at $N=10$ corresponds to Plato's polyhedron.
The numerous highly symmetric structures including Plato's polyhedron are labeled as TETRA, OCTA, ICO, CUBO, and DECA,
and they respectively represent a tetrahedron, octahedron, icosahedron, cuboctahedron, and decahedron, respectively.
%
As shown in FIGS.2 and 3 the three structures composed of blue, red, and silver spheres
represent the structures of Na$_N$, Cu$_N$, and Ag$_N$, respectively
These structures are shown to emphasize the relatively large structural differences from the relation of similarity.
%
%

%
The following conclusions are obtained from the most and meta-stable structures of the clusters.
The optimized structures of the Na$_N$, Cu$_N$, and Ag$_N$ clusters are almost identical if the most
and meta-stable structures are simultaneously considered.
In particular, the most stable structure of the three clusters for $N\leq 22$ can be identified with high probability.
For $N=34$ and $38$, all of the most stable structures of the three clusters are different.
For $N=40$, the most stable structures of alkali and noble metal clusters clearly differ.
For $N=55$, all of the three different clusters favor structural type I  (icosahedral structure) for their most stable structure.
The 2nd and 3rd stable structures at $N=55$ include an atomic vacancy in an icosahedral flame \cite{Itoh-2}. 
For all of the three different metal clusters, these meta-stable structures are more stable
than the higher symmetric structures such as a decahedron and cuboctahedron.
Generally, it is difficult to define an atomic vacancy in a cluster because it is a system without periodicity in the structure.
However, an icosahedral structure can be considered to be a structure that is cleaved from a quasicrystal.
In fact, $N=55$ is the minimum number that shows an atomic vacancy stability in the cluster structure.
For $N\geq 55$, with increasing $N$, structures of the clusters approach the periodic bulk crystal structure.
Therefore, it is expected that the probability of vacancy formation will increase in finite cluster systems.
%
%

An overview of the structural type transition of the most stable structures
in Na$_N$, Cu$_N$, and Ag$_N$ clusters with increasing $N$ is described below.
The structures of Na$_N$, Cu$_N$, and Ag$_N$ clusters are stable in a linear structure at $N=2$,
planar structure at $3\leq N \leq 5$ (Na), $6$ (Cu, and Ag),
opened structure at $6$ (Na), $7$ (Cu and Ag) $\leq N \leq 15$ (Na and Cu), $16$ (Ag), and closed structure at $N \geq 16$ (Na and Cu), $17$ (Ag).
Systems comprising for or more atoms can possibly assume three-dimensional structures such as a tetrahedron and trigonal pyramid.
However, for all of the three different metal clusters, the most stable structures are not three-dimensional at $4\leq N \leq 5$ (Na), $6$ (Cu, and Ag).
At $N=4$, the 3 dimensional structures are not stable for these systems. Instead, a rhombus structure is stable at $N=4$.
A three-dimensional structure such as a tetrahedron is realized in other metal clusters such as Mg$_4$ \cite{Mg}.

%
The stable cluster structures obtained in this study can be roughly understood from knowledge of the electronic states using the spherical jellium model.
The main difference between the first principles model used in this study and the spherical jellium model
is the manner in which the atomic configuration in the cluster is treated.
In the spherical jellium model, only the outermost valence electrons in the each atom composing the cluster are considered as valence electrons.
The other electrons and nuclei are unified into an uniform positive charge distribution with a spherical surface boundary.
Therefore, the valence electrons in the spherical jellium model reside in the central force field.
As a result, each degenerated valence electronic state is specified with a monoangular momentum: $S, P, D, F, ...$, in a manner
similar to the case of electrons in an atom and protons and neutrons in a nucleus.
The stable structures of the metal cluster obtained in this study can be roughly considered
as those satisfying the cluster shapes followed by the valence electronic density distribution
that originates from the occupied orbitals in the spherical jellium model.

\subsection{$N$ dependence of the averaged nearest neighbor distance and coordination number}
Figure 4 (a) shows the averaged nearest neighbor distance (ANND)
of the most stable structures of Na$_N$, Cu$_N$, and Ag$_N$ clusters for $2\leq N \leq 75$.
ANND is defined as the sum of the nearest neighbor atomic distances divided by the number of bonds in a cluster.
As shown in the figure, the ANND value of the cluster reaches approximately 90$\%$ of the bulk value at $N=20$ for all of the three different clusters.
However, the ANND does not converge to the bulk value at $N=75$, and the values are approximately $98\%$ of the bulk ones.
Over the entire range of $N$, the ratio of ANND among the three different clusters agrees well with that of the bulk crystals.

Figure 4 (b) shows the averaged coordination number (ACN) of the most stable structures
of Na$_N$, Cu$_N$, and Ag$_N$ clusters over the range of $2\leq N \leq 75$.
ACN is defined as the sum of the nearest neighbor coordinated atomic number for all atoms composing the cluster divided by $N$.
The $N$ dependence of ACN for the three different clusters is similar;
it should be noted that the ACN values do not reach 9 (75$\%$ of the bulk value of 12) even at $N=75$.
This is attributable to the existence of a surface in the cluster. 
\subsection{Difference between planar structures of sodium, copper, and silver clusters}
As described in section III. A., the most and meta-stable structures of the three different metal clusters are quite similar.
Here, the cluster structures having the same structural type are compared quantitatively using structure III at $N=7$ as an example.
FIG. 5 shows the details of the structures,
represented by the Cu$_7$ structure as an example.
The three values represent the angle or interatomic distance of Na$_7$, Cu$_7$, and Ag$_7$, respectively.
These values are expressed in degrees or angstroms, respectively.
The blue and green values in parentheses represent the relative angle ratios: angle (distance)$_{Na_7}$/angle(distance)$_{Cu_7}$
and angle(distance)$_{Ag_7}$/angle(distance)$_{Cu_7}$, respectively.

Structure III in Na$_{7}$ belongs to the $C_{2h}$ point group, while in Cu$_{7}$ and Ag$_{7}$, it belongs to the $D_{2h}$ point group;
all of these structures are 3rd stable structures.
These structures have a Jahn-Teller deformation \cite{JT} in the equilateral hexagon in the $D_{6h}$ point group.
As expected, the relative angle ratio of Ag$_7$ is closer to 1 than that of Na$_7$.
This is a typical example that indicates the stronger similarity between the two noble metal clusters.
However, there are several exceptions in clusters having a larger value of $N$.
This may occur due to the increase in the number of degrees of freedom in the atomic positions.
Therefore, it is not possible to present an oversimplified picture of the degree of similarity in these cluster structures.
%
\subsection{$N$ dependence of the averaged atomic density}
The averaged atomic volume of cluster $v(N)$ is defined as
\begin{equation}
v(N)\equiv \frac{4}{3}\pi \langle R\rangle^{3}/N.
\end{equation}
Here, $\langle R\rangle$ indicates the an averaged distance between each atomic coordination
$\bm{R} _{i}$ and center of mass $\bm{R} _{CM}$ in a cluster.
In other words, $\langle R\rangle$ is given as
\begin{equation}
\langle R\rangle=\frac{1}{N}\sum_{i=1}^{N}\bigl | \bm{R} _{i}-\bm{R} _{CM}\bigr |.
\end{equation}
To compare the $N$ dependence of $v(N)$ for the different metal clusters, each $v(N)$ is normalized to that of a dimer. i.e., $v(2)$.

Figure 6 (a) shows the $N$ dependence of a normalized volume for the most stable structures of Na$_N$, Cu$_N$, and Ag$_N$ for $2\leq N\leq 22$.
The $N$ dependence of the normalized volume $v(N)/v(2)$ is very similar in the three metal clusters.
The trends of the change in the $N$ dependence at transition sizes $N$ between different structural types are indicated using the symbols L, P, O, and C.
$v(N)/v(2)$ decreases significantly at $N$ from P to O. Further, significant decreases are observed at $N$ from O to C.
Figure 6 (b) shows the volumes of the most stable structures in the range of $15 \leq N \leq 75$.
Here, the scale of $v(N)/v(2)$ is expanded to show the change clearly.
In general, for $N\geq 20$, $v(N)/v(2)$ increases monotonically with $N$ for all of the three metal clusters.
However, it should be noted that the values decrease significantly with increasing $N$
between Cu$_{40}$ (Ag$_{40}$) and Cu$_{55}$ (Ag$_{55}$), unlike the case of Na$_{40}$ and Na$_{55}$.
This can be attributed to the effect of \textit{d} electrons
in that they may shrink the interatomic distances of the quasi-spherical structure-an icosahedron which is the most stable structure at $N=55$.

Figure 6 (c) shows $v(N)/v(2)$ of the most stable structure (icosahedron) of Na$_N$
for $N=55, 147$, and $309$ obtained in our previous study \cite{Itoh-2}.
Here, $v(N)/v(2)$ increases monotonically with $N$.
FIG. 6 (d) shows the melting points $T_{m}$ of the Na$_N$ cluster at $N=55, 147$, and $309$,
as observed by Haberland \textit{et al.} \cite{Haberland-2005}, to discuss the relationship with cluster volumes.
$T_{m}$ decreases monotonically with increasing $N$.
This $N$ dependence of $T_{m}$ is in contrast to the trend observed in the case of the cluster volume.
Aguado \textit{et al.} showed that the volume $v(N)$ and $T_{m}$ of the most stable structures of the Na$_N$ cluster
exhibited an opposite trend for $N\geq 55$ from the molecular dynamics calculations based on DFT-LDA \cite{Aguado}.
This conclusion supports the existence of a relationship between the $N$ dependence of $v(N)/v(2)$ as evaluated by us
and that of the $T_{m}$ observed by Habarland \textit{et al}.
Further, the more precise DFT-LDA molecular dynamics calculations predict that $T_{m}$ of Na$_{40}$ is higher than that of Na$_{55}$ \cite{Lee}.
$T_{m}$ of Na$_{40}$ has not yet been observed experimentally.
Our calculation results of $v(N)/v(2)$ do not contradict
with the result of the DFT-LDA molecular dynamics study \cite{Lee} with regard to the relationship between Na$_{40}$, and Na$_{55}$.

If the conclusion about the relation between the cluster $v(N)$ or $v(N)/v(2)$ and $T_{m}$ in Na$_N$
holds for Cu$_N$ and Ag$_N$ for $N\leq 55$,
it is expected that $T_{m}$ of Cu$_{40}$ and Ag$_{40}$ are lower than that of Cu$_{55}$ and Ag$_{55}$, respectively.
For $N\leq 22$, the volume of the most stable cluster changes significantly with 
the structural type transitions (L$\rightarrow$ P$\rightarrow$ O$\rightarrow$ C) described in section III. A.
Therefore, for $N\leq 22$, the $N$ dependence of the cluster $T_{m}$ is expected to reflect this change.
\section{$N$ dependence of the calculated cluster energy for $2\leq N \leq 75$ and $\infty$}
\subsection{Overall aspects of the cluster binding energy}
As described in section III. A., the most stable structures of the three metal clusters were searched over the range $2\leq N \leq 75$
and the ground state cluster energies $E(N)$ were evaluated. In general, it was observed that $E(N)/N$ of the metal clusters increased with $N$ 
and approached $E(\infty)/\infty$, which corresponds to the value of the bulk cohesive energy.
Interestingly, $E(N)$ exhibited higher or lower values at a particular value of $N$.
To investigate the aspects of $E(N)$ in datail, the $N$ dependence of the difference between $E(N)$
and the liquid drop model (LDM) \cite{LDM} average $\langle E(N)\rangle$, defined as
\begin{equation}
\delta E(N)\equiv\langle E(N)\rangle-E(N),
\end{equation}
is evaluated.
Here, $\langle E(N)\rangle$ is expressed by a linear combination of three terms as shown below;
\begin{equation}
\langle E(N)\rangle= a_{v}N+a_{s}N^{\frac{2}{3}}+a_{c}N^{\frac{1}{3}}.
\end{equation}
The first, second, and third terms denote the volume, surface, and curvature energy, respectively.
The fitting parameters $a_{v}$, $a_{s}$, and a$_{c}$ were determined as described below.
Here, the averaged cluster binding energy per atom $\langle E_{b}(N)\rangle/N$, defined as
\begin{equation}
\langle E_{b}(N)\rangle/N\equiv E(1)-\langle E(N)\rangle/N
\end{equation}
must be calculated to evaluate $\delta E(N)$.

$a_{v}$ was uniquely determined from the cohesive energy of the bulk crystal that corresponds to $\langle E_{b}(N)\rangle/N$ at $N=\infty$.
$a_{s}$ and $a_{c}$ were determined as coefficients of the fitting curve $\langle E_{b}(N)\rangle/N$
to the cluster binding energy $E _{b}(N)/N$ over the range of $2\leq N\leq 75$
by applying the least squares method and minimizing those values under the threshold $\langle E_{b}(1)\rangle=0$.
FIG. 7 shows the cluster binding energy per atom $E_{b}(N)/N$ and the LDM average $\langle E_{b}(N)\rangle/N$ as functions of $N^{-\frac{1}{3}}$.
$E_{b}(N)/N$ and the average $\langle E_{b}(N)\rangle/N$ of Na$_N$, Cu$_N$, and Ag$_N$ at $N=2-75$ and $\infty$ are presented.
$N^{-\frac{1}{3}}=0, 0.237$ and $0.794$ correspond to $N=\infty, 75$, and $2$, respectively.

We now consider the difference between $E_{b}(N)/N$ and the LDM average $\langle E_{b}(N)\rangle/N$.
With an increasing in $N$ from $2$ to $75$ and then $\infty$, a significant similarity is observed in the difference
among the Na$_N$, Cu$_N$, and Ag$_N$ clusters in which each element has one \textit{s} electron in the outermost shell.
For the sake of comparison, the experimental bulk cohesive energy and binding energy of a dimer \cite{CRC, Kittel, expt} are also shown in FIG. 7.
The values of the bulk cohesive energy and binding energy obtained through our calculations and experiments are in good agreement.
However, in an Ag bulk crystal, the calculated cohesive energy does not agree well with the experimental value.
This disagreement may be attributable to the incompleteness of the description of the Ag ($Z=47$) atom.
As reported in previous studies \cite{Ag}, the disagreement originates from the treatment method of the relativistic effect through the pseudopotential,
the exchange-correlation energy functional - GGA, and the basis set for the electrons - plane waves.
Although an improvement in the description is desired, we expect that the qualitative feature of the growth behavior
of an Ag cluster from an atom to bulk shown in this study will not change.

\subsection{Detailed analysis of the $N$ dependence of cluster binding energy}
%
%
In section IV. A., a strong similarity is pointed out for Na$_N$, Cu$_N$, and Ag$_N$ clusters
in the $N$ dependence of the difference between the cluster binding energy per atom-$E_{b}(N)/N$- and the LDM average-$\langle E_{b}(N)\rangle/N$.
In this section, we analyze this similarity in detail.
FIG. 7 shows that the difference for each of the three clusters approaches 0 with increasing $N$.

However, it is difficult to compare them for large value of $N$.
Therefore, the $N$ dependence of the $N$ multiplied values
$\delta E(N)\equiv -E(N)+\langle E(N)\rangle$
are noted for the detailed analysis.

Figure 8 (a) and (b) show the $N$ dependence of $\delta E(N)$ for the three metal clusters
over the ranges $1\leq N \leq 22$ and $15\leq N \leq 75$, respectively.
In FIG. 8 (a), a strong similarity is observed in the $N$ dependence of $\delta E(N)$ for $1\leq N \leq 22$.
These values indicate an odd-even alternation in $N$.
In other words, $\delta E(N)$ increases from an odd $N$ to the next even $N$, and then decreases from an even $N$ to the next odd $N$.
In many cases, $\delta E(N)$ is positive at even $N$ and negative at odd $N$.
In FIG. 8 (b), a significant similarity is observed in the $N$ dependence of $\delta E(N)$ for $15\leq N \leq 75$.
$\delta E(N)$ is larger at $N=34$ (Na), $55$, and $58$ and smaller at $N=40$ (Cu and Ag), $68$, $70$, $71$, and $75$
are compared to those at the neighborhood $N$ considered in this study.
The amplitudes of $\delta E(N)$ for Cu$_N$ and Ag$_N$ are much larger than that for Na$_N$.
Further, the values of Cu$_N$ and Ag$_N$ are closer. This is attributable to the effect of \textit{d} electrons in noble metal clusters.
\subsection{Definition of the cluster magic number and identification}
%
In many previous studies of clusters, an $N$ value that gives a special cluster stability has been frequently called as the magic number.
To understand the $N$-dependent system stability,
it is necessary to evaluate the magic number based on first principles calculations,
because it can be a standard magic number.
However, the magic number has not yet been defined.
Therefore, to understand the $N$-dependent system stability, we must first define the magic numbers.

First, the curvature of the cluster energy for $N$-$\Delta _{2}E(N)$, defined as
\begin{equation}
\Delta _{2}E(N)\equiv E(N+1) + E(N-1) - 2E(N),
\end{equation}
is discussed.
If $\Delta _{2}E(N)$ exhibits a positive peak at $N$, $N$ may be a magic number because $E(N)$ might be a local minimum for $N$.
FIG. 8 (c) shows the $N$ dependence of $\Delta _{2}E(N)$ for the three metal clusters.
Here, the magic numbers of the three clusters can be identified as $N=2, 4, 6$ (Na, Ag), $8, 10, 12, 14, 18,$ and $20$. 
For these values of $N$, an odd-even alternation is clearly observed.
All of these values of $N$ are even numbers and it is natural to consider that
the odd-even alternation originates from the shell closing of each electronic state in the clusters.
In addition, the $N$ dependences of the peak intensities are similar in the three clusters,
although the absolute values of the peaks are larger in the noble metal clusters than in the alkali metal cluster for most values of $N$.
As in the case of $\delta E(N)$ described in IV. B., the values of $\Delta _{2}E(N)$ are closer in the noble metal clusters.

Although it is possible to derive other conclusions from the absolute values of $\Delta _{2}E(N)$,
only $\Delta _{2}E(N)$ may not be a sufficient criterion for the evaluation of magic numbers.
In general, additional information is required for the identification.

$\delta E(N)$ can be a criterion for identifying magic numbers because of its definition:
the energy difference between a real cluster and the continuously averaged energy model, LDM.
The magic numbers $N$ should satisfy the condition $\delta E(N)\geq 0$.
This condition is satisfied for $N=2, 4$ (Ag), $6, 7, 8, 10, 12$ (Cu), $14, 17$ (Cu), $18, 19, 20$, and $21$ (Na).

If only $\Delta _{2}E(N)$ and $\delta E(N)$ are considered, magic numbers $N$ can be defined as
\begin{equation}
\Delta _{2}E(N) \geq 0, \delta E(N) \geq 0.
\end{equation}
If this definition is employed for the magic numbers in Na$_N$, Cu$_N$, and Ag$_N$ clusters for $N\leq 21$,
the common magic numbers $N=2, 4$ (Ag), $6, 8, 10, 12, 14, 18$, and $20$ are identified.

In this study, $\Delta _{2}E(N)$ for $N\geq 34$ is not evaluated.
For $N\geq 34$, magic numbers are identified only from $\delta E(N)$.
For these values of $N$, we consider those that satisfy
\begin{equation}
\delta E(N)\geq 0
\end{equation}
to be magic numbers.
Therefore, $N=34$ (Na), $38$ (Na), $40$ (Na), $55$, and $58$ are identified as magic numbers.
In this definition, the odd number $N=55$ is also considered to be a magic number.

In our first principles calculations, the common feature of the magic numbers in Na$_N$, Cu$_N$, and Ag$_N$
can be attributed to the delocalized \textit{s} valence electrons.
The localized \textit{d} electrons in the noble metal clusters increase the stability difference between magic and not-magic clusters.

\subsection{$N$-dependent stability of Na$_N$ clusters from the experiment and total energy calculations}
The $N$ dependence of the peak intensity $I(N)$ in the experimentally observed abundance spectra,
and the cluster energy $E(N)$ evaluated from the first principles calculations have not yet been compared quantitatively.
This comparison is important for understanding the magic number observed in the experiment. 
On the basis of several assumptions, semi-quantitative comparison method \cite{Knight-deHeer, deHeer-RMP} has already been developed
for an experimental result of Na$_N$ \cite{Knight} and for the theoretical total energy calculations.
%
Although experimental results for Cu$_N$ and Ag$_N$ \cite{Katakuse-1, Katakuse-2} are available,
they cannot be easily compared with our calculation results quantitatively.
Therefore, in this section, we only focus on the magic numbers of Na$_N$.

With regard to experiments with Na$_N$, only the result of Knight \textit{et al}. \cite{Knight} is discussed.
As shown in other experiments with Na$_N$ such as those by Bj$\o$rnholm \textit{et al}. \cite{Bjornholm},
Rabinovitch \textit{et al}. \cite{Rabinovitch}, etc., main peaks were also observed at $N=8, 20, 40, 58$, and $92$.
Therefore, without significant improvements in the experimental method, these abundance spectra and magic numbers appear to remain unchanged.
Obviously, if the experimental results remain unchanged, the theory itself must be reconsidered.

Here, the curvature of $E(N)$ evaluated from our calculations and the peak intensity $I(N)$ in the abundance spectra
were semi-quantitatively compared based on the method described below.
The abundance spectra of Na$_N$ clusters reported by Knight \textit{et al}. are considered to be attributed as follows.
First, Na atomic vapor is formed by heating Na bulk solid in an Ar-gas-filled closed space in the experimental apparatus \cite{Knight-Ar}.
Then, Na$_N$ clusters are formed by the adiabatic cooling of the mixture after passing through the skimmer in the apparatus.
Then, $N$-distributed neutral charged Na$_N$ clusters were assumed to be obtained in a thermal equilibrium condition.
If the assumption for the Na$_N$ distribution is true, the relationship 
\begin{equation}
\rho(N)=A exp \bigg(-\frac{E(N)}{k_{B}T}\bigg)
\end{equation}
must be satisfied.
Here, the $\rho(N)$, $A$, $k_{B}$, and $T$ denote the number density of neutral charged Na$_N$,
a constant, Boltzmann constant, and the absolute temperature, respectively.
These clusters were ionized by light and then accelerated by an electric field for mass selection.
In these steps, the $N$-distribution of $\rho(N)$ is assumed to remain unchanged.
Finally, these clusters were detected and the $N$-distribution of the neutral charged clusters was observed using the detector.
Therefore, finally,
\begin{equation}
I(N)=B\rho(N)
\end{equation}
holds. Here, $B$ is a constant. From equations (6), (9), and (10), the following equation is derived.
\begin{equation}
\Delta _{2}E(N)=k_{B}Tln\frac{ I(N)^{2} }{ I(N+1)I(N-1) }
\end{equation}
Namely, $\Delta _{2}E(N)$ can be evaluated from $I(N)$ of the experimental abundance spectra.

Figures 8 (d) and (e) show the $N$ dependence of $\Delta _{2}E(N)$ evaluated experimentally by Knight \textit{et al.}-$\Delta _{2}E_{Expt.}(N)$
and by the first principles calculations in this study-$\Delta _{2}E_{FP-DFT}(N)$ for $N\leq 22$ and $15\leq N\leq 75$, respectively.
For the latter discussion, $\delta E(N)$ evaluated from the first principles calculations in this study, $\delta E_{FP-DFT}(N)$,
and from the DFT-based spherical jellium model (SJM) calculations by Genzken \textit{et al.} \cite{Genzken}, $\delta E_{SJM-DFT}(N)$,
are also shown in these figures.

We first compare the $N$ dependence of $\Delta _{2}E_{Expt.}(N)$ and $\Delta _{2}E_{FP-DFT}(N)$.
Generally, there is a good agreement between them.
However, contrasting trends are observed in two rows of $N$: (1) $N=5, 6$ and $7$ and (2) $N=17$ and $18$.
The peak intensity of $\Delta _{2}E_{Expt.}(N)$ and $\Delta _{2}E_{FP-DFT}(N)$ is relatively strong at $N=8$ and $20$.
For $N\geq 34$, $\Delta _{2}E_{FP-DFT}(N)$ is not evaluated in this study.

As shown in Figures 8 (d) and (e), there is a similarity in the $N$ dependence between $\Delta _{2}E_{FP-DFT}(N)$ and $\delta E_{FP-DFT}(N)$.
Considering this similarity, instead of $\Delta _{2}E_{FP-DFT}(N)$, $\delta E_{FP-DFT}(N)$ can be considered
as a comparative value to $\Delta _{2}E_{Expt.}(N)$.

In a manner similar to $\Delta _{2}E_{FP-DFT}(N)$, $\delta E_{FP-DFT}(N)$ exhibits good agreement
with $\Delta _{2}E_{Expt.}(N)$ with regard to the $N$ dependence.
However, in a manner similar to the case of $\Delta _{2}E_{FP-DFT}(N)$, $\delta E_{FP-DFT}(N)$ exhibits
a different $N$ dependence as compared to $\Delta _{2}E_{Expt.}(N)$ in two rows of $N$: (1) $N=5$ and $6$ and (2) $N=17$ and $18$.
Here, it should be noted that the order of peak intensity at $N=8$ and $20$ of $\delta E_{FP-DFT}(N)$
exhibits better agreement with $\Delta _{2}E_{Expt.}(N)$ than with $\Delta _{2}E_{FP-DFT}(N)$.
Based on these results, $\delta E_{FP-DFT}(N)$ is compared with $\Delta _{2}E_{Expt.}(N)$ for $N\geq 34$.

In $\Delta _{2}E_{Expt.}(N)$, distinctive peaks are formed at $N=40$ and $58$.
Similar to the case of $\Delta _{2}E_{Expt.}(N)$, $\delta E_{FP-DFT}(N)$ is relatively large at $N=58$. 
However, $\delta E_{FP-DFT}(N)$ at $N=40$ is not particularly so distinctive.
Although $\Delta _{2}E_{Expt.}(N)$ is remarkably distinctive at $N=40$,
the value of $\delta E_{FP-DFT}(40)$ is similar to that of $\delta E_{FP-DFT}(38)$.
Further, for $N=34$ and $55$, $\delta E_{FP-DFT}(N)$ exhibits distinctive peaks
and these values are larger than $\delta E_{FP-DFT}(40)$ and $\delta E_{FP-DFT}(58)$.
In addition, for $N=68, 70, 71$, and $75$, $\delta E_{FP-DFT}(N)$ has negative values,
and these features are not exhibited by $\Delta _{2}E_{Expt.}(N)$.

As shown here for Na$_N$, although the $N$ dependence of $\delta E_{FP-DFT}(N)$
exhibits good agreement with that of $\Delta _{2}E_{Expt.}(N)$ in the range of $N\leq 22$,
the same is not necessarily true in the range of $N\geq 34$.
This point is discussed in the next subsection.

\subsection{Improvement of the magic number description for Na$_N$ clusters from first principles calculations}

We evaluate the source of this disagreement in the $N$ dependence of $\delta E(N) (\Delta _{2}E(N))$
obtained from theoretical total energy calculations and $\Delta _{2}E(N)$ obtained from the experiment.
In this subsection, we review the $N$-dependent system stability of Na$_N$ by the improvement of theoretical models.

As shown in Figures 8 (d) and (e), $\delta E_{SJM-DFT}(N)$ obtained by Genzken \textit{et al.} \cite{Genzken}
exhibits an $N$ dependence similar to that of $\delta E_{FP-DFT}(N)$ for $2\leq N\leq 58$.
However, the $N$ dependence of $\delta E_{FP-DFT}(N)$
exhibits a better agreement with that of $\Delta _{2}E _{Expt.}(N)$ for $N=34, 40$, and $58$ as compared to that of $\delta E_{SJM-DFT}(N)$.
This resut suggests that the consideration of the explicit ionic configuration in the total energy calculation is significant
to realize a better agreement with the experiment with regard to the $N$-dependent system stability of Na$_N$.

Figures 8 (f) and (g) show the $N$ dependence of $\Delta _{2}E(N)$ for Na$_N$ evaluated from various DFT-based models such as
the spherical jellium model (SJM-DFT) by Chou \textit{et al.} \cite{Chou},
spheroidal jellium model (S'JM-DFT) by Ekardt \textit{et al.} \cite{Ekardt-1988},
ultimate jellium model (UJM-DFT) by Koskinen \textit{et al.} \cite{UJM},
and the first principles model (FP-DFT) in this study for $1\leq N\leq 22$ and $15\leq N\leq 75$, respectively.
These $\Delta _{2}E(N)$ values are denoted as $\Delta _{2}E_{SJM-DFT}(N)$, $\Delta _{2}E_{S'JM-DFT}(N)$,
$\Delta _{2}E_{UJM-DFT}$, and $\Delta _{2}E_{FP-DFT}(N)$, respectively.
The various jellium models are differ from the first principles calculations in the treatment of the ionic configuration and the exchange-correlation energy.

As shown in Figures 8 (f) and (g), all of the calculation results exhibit distinctive peaks at $N=8, 18$, and $20$.
In these figures, remarkable differences are obtained in the number of peaks between $\Delta _{2}E(N)$
of the spherical models, SJM-DFT, and non-spherical models, S'JM-DFT, UJM-DFT, and FP-DFT.
From these figures, it is apparent that the improvement in the $N$ dependence in $\Delta _{2}E(N)$
from the jellium models to the first principles model that explicitly treats the ionic configuration is smaller as
compared to that from the spherical models to the non-spherical models.

In a manner similar to the case of $\delta E_{FP-DFT}(N)$ and $\Delta _{2}E_{FP-DFT}(N)$ for $1\leq N \leq 22$,
as shown in Figures 8 (e) and (g), the $N$ dependence of $\Delta _{2}E_{SJM-DFT}(N)$
is similar to that of $\delta E_{SJM-DFT}(N)$ over the range of $2\leq N \leq 58$.
Here, it is assumed that there is a similarity
between the $N$ dependence of $\delta E_{FP-DFT}(N)$ and that of $\Delta _{2}E_{FP-DFT}(N)$ for $34\leq N\leq 58$.

As shown in Figure 8 (e), $\Delta _{2}E_{SJM-DFT}(40)$ ($\Delta _{2}E_{S'JM-DFT}(40)$) is positive.
However, it is smaller than $\Delta _{2}E_{SJM-DFT}(20)$, $\Delta _{2}E_{SJM-DFT}(34)$, and $\Delta _{2}E_{SJM-DFT}(58)$
($\Delta _{2}E_{S'JM-DFT}(20)$, $\Delta _{2}E_{S'JM-DFT}(34)$, and $\Delta _{2}E_{S'JM-DFT}(58)$).
On the other hand, as shown in FIG. 8(g), $\delta E_{SJM-DFT}(N)$ for $N=40$ is negative.
Therefore, based on the definition of the magic number given in equation (7), $N=40$ is not a magic number in SJM-DFT.
However, $N=40$ may be a magic number in S'JM-DFT and FP-DFT although $\delta E_{FP-DFT}(40)$ has a small positive value.
Although the magic feature at $N=40$ is expected to be weak in S'JM-DFT and FP-DFT,
it is also expected that the consideration of the deviation from central force field in the total energy calculation is significant
for obtaining a better agreement in $N$-dependent system stability with the experiment.

Here, it should be noted that $\Delta _{2}E_{SJM-DFT}(68)$ is nearly 0 eV.
As shown in FIG. 8 (e), there is a large difference between $\Delta _{2}E_{Expt.}(68) \sim$ 0 eV and $\delta E_{FP-DFT}(68) \sim$ $-0.5$ eV.
From the similarity between the $N$ dependence of $\delta E(N)$ and $\Delta _{2}E(N)$,
$\Delta _{2}E_{FP-DFT}(68)$ is expected to differ significantly from $\Delta _{2}E_{Expt.}(68)$ as compared to $\Delta _{2}E_{SJM-DFT}(68)$.
Therefore, it can be said that the $N$ dependences of $\Delta _{2}E_{FP-DFT}(N)$ and $\delta E_{FP-DFT}(N)$ are not necessarily
more similar to that of $\Delta _{2}E_{Expt.}(N)$ as compared to those of $\Delta _{2}E_{SJM-DFT}(N)$ and $\delta E_{SJM-DFT}(N)$.
%

We now consider the possibilities for reducing the difference between the theoretically and experimentally obtained $N$-dependent system stability.
First, we discuss the possibility of improving has the magic number description based on the evaluation of the ground state energy.
In the case of Na$_N$, it has already been shown that the $N$ dependence of the experimental abundance peak at $N=5, 6$, and $7$
cannot be expressed even in terms of $\Delta _{2}E(N)$ of the configuration interaction calculations \cite{Koutecky-1, Koutecky-2}.
Therefore, it is expected that the improvement of the ground state energy evaluation will never lead to and improvement in the agreement
for small values of $N$.

Thus for, several researchers have shown stable structures of Na$_{40}$ \cite{Na40, Lee, Ghazi}.
%
%
In previous studies, a nearly spherical structure with high symmetry has not yet been obtained as the most stable structure.
Instead of such a structure, we have found a structure with low symmetry (III) that is 0.320 eV more stable
than structure (II), which is similar to the most stable structure found thus far \cite{Lee}.
Since the energy differences between these structures are small, the most stable structure, III,
did not exhibit a strong peak for $N=40$ in $\delta E_{FP-DFT}(N)$.
Therefore, we expect that $\delta E_{FP-DFT}(40)$ will not realize a strong magic feature even if a more stable structure is found in Na$_{40}$.

Within SJM, the $N$ dependence of the energy per valence electron for Na$_N$ was evaluated
from a more precise first principles calculation method called the diffusion Monte Carlo (DMC) calculation \cite{Tao}.
A weak magic feature was obtained for $N=40$ as a dip in the total energy vs. $N$ curve.
From the result a weak magic feature is expected for $N=40$
in the $N$ dependences of $\Delta _{2}E_{SJM-DMC}(N)$ and $\delta E_{SJM-DMC}(N)$,
in a manner similar to the cases of SJM-DFT and FP-DFT.
However, by considering the explicit geometrical structure in DMC,
the magic feature for $N=40$ will be strengthened as compared to that for $N=34$ and $58$.

Second, we discuss the possibility of  realizing improvements by a more proper evaluation of the system stability.
In the experiments, the temperature in the apparatus is expected to be related to the final results.
However, in the cluster study based on the total energy calculations, the temperature of the system is 0 K.
If the consideration of the temperature is critical for the $N$-dependent system stability,
instead of the internal energy $E(N)$, the free energy $F(N)$ must be considered. 
Further, the most stable and meta-stable structures of Na$_N$ must both be considered properly for the $N$-dependent system stability.
To consider the contribution of meta-stable structures, information about the potential energy surface or free energy landscape is required.

Finally, we discuss the possibility of realizing improvements from other viewpoints.
In the case of Na$_N$, various shell models such as the Woods-Saxon type \cite{Knight, Nishioka} and harmonic oscillator type \cite{Clemenger}
that neglect the Coulomb interactions between electrons exhibit a magic feature at $N=40$
in the $N$ dependence of the shell correction energy and the curvature for the sum of the electronic eigenvalues.
These $N$ dependences are strongly affected by the $N$ dependence of the energy gap
between the highest occupied molecular orbital (HOMO) and the lowest unoccupied molecular orbital (LUMO), i.e., the HOMO-LUMO gap.
As shown in FIG. 9, within FP-DFT, the HOMO-LUMO gap for the most stable structure of Na$_{40}$ is larger than that of Na$_{34}$.
The absolute value of the HOMO-LUMO gap may be larger if the quasi-particle energy evaluated from GWA is considered \cite{Yoshizaki}.
The improvement is significant if $I(N)$ in the abundance spectra of Na$_N$ is a value that is more strongly related
to the shell correction energy and the curvature of the sum of electronic eigenvalues as compared
to $\Delta _{2}E(N)$ or $\delta E(N)$ evaluated from $E(N)$.

As described in this section, by a more proper treatment of the system within the total energy calculations,
discrepancies in the description of the $N$-dependent system stability with the experimental results in the description of can be reduced.
%
%
However, this is difficult even in the case of the simplest metal cluster Na$_N$ shown here.
For discussing the $N$-dependent cluster stability, various problems must first be solved.
However, the magic number for the ground state, which avoids empirical parameter, is significant
because it can be a standard to understand the $N$-dependent system stability.
Further, for practical reasons, the evaluation method based on first principles calculations
is expected to retain the value in predictions of the experimental values of magic numbers.

\section{Electronic Structures}
%
\subsection{Evolution of the electronic structure from atom to bulk}
As described before, remarkable similarities are observed in the $N$ dependence of the most and meta-stable structures
and in the ground state energies of Na$_N$, Cu$_N$, and Ag$_N$ clusters.
These similarities are attributed to the $N$ dependence of the electronic structure.
In this section, $N$-dependent electronic structures for the most stable structures of each metal cluster are discussed.

Figures 9 (a), (b), and (c) show the Kohn-Sham energy spectra
for the most stable structures of the three metal clusters over the range of $1\leq N \leq 75$ and $\infty$ (bulk).
The density of states of the bulk crystal obtained from the band calculation is shown
in the right-hand side space of each figure for the sake of comparison.
Each Fermi level of the bulk crystal is set to the next HOMO-LUMO gap for the $N=75$ cluster.
For all of the three figures, the space between HOMO and LUMO is colored blue.
The red lines represent the occupied and unoccupied electronic energy levels.

For most values of $N$, the HOMO-LUMO gaps are small at odd values of $N$ and large at even values of $N$,
which results in the odd-even alternation of the HOMO-LUMO gaps.
However, the HOMO-LUMO gap is large for several odd values of $N$ and small for even values of $N$.
It should be noted that the odd-even alternation of the HOMO-LUMO gaps in Cu$_N$ and Ag$_N$
are significantly larger than that of Na$_N$ for the benefit of \textit{d} electrons.
Further, the similarity in the odd-even alternation is more distinctive between the noble metal clusters.
Generally, the HOMO-LUMO gap of each metal cluster converges to the Fermi level of the bulk crystal with increasing $N$.

For all of the metal clusters, with increasing $N$, the feature of each bulk energy band gradually appears in the electronic states.
The energy width between the bottom and top of the occupied energy levels almost converges to that of the bulk crystal at $N=75$.
In the case of Na$_N$, over a wide energy range, the states are characterized by \textit{s} and \textit{p}. 
In Cu$_N$ and Ag$_N$, in addition to the \textit{s}, and \textit{p} characterized states
similar to those shown in Na$_N$, energetically localized \textit{d} band type states also appeared.
It should be noted that the features of localized \textit{d} band in the bulk crystals already appeared
at a rather small value of $N$ in the width of the \textit{d} states and the energetical distance from the top of the \textit{d} states to the Fermi level.
With increasing $N$, the strongly \textit{d} characterized states gradually expand
and almost converge to the width of the \textit{d} band in each bulk crystal at $N=75$.

For the same type of geometrical structures as those described in section II. A.,
distinctive simiralities were also observed in the underwent splitting manner in which electronic states
that are strongly characterized by \textit{s} and \textit{p}.
As shown in FIG. 9, the energetical distances for the adjacent states strongly characterized
by \textit{s} and \textit{p} are larger in Na$_N$ than those of Cu$_N$ and Ag$_N$.
This relationship can be roughly understood from the relationship between the energetical distance for the adjacent states
and the width of a well in the quantum well model, as described in the previous paper \cite{Itoh-1}.
In this case, the \textit{s} and \textit{p} characterized electrons correspond to the quantum in the well.
As shown in FIGS. 4 (a), 5, and 6 (a) and (b), for the existence of spatially and enegetically localized \textit{d} electrons,
closer interatomic distances are obtained in noble metal clusters as compared to those in alkali metal clusters.
Namely, the width of the well is narrower in the noble metal clusters.
As a result, larger energetical distances are realized in the noble metal clusters.

The stability of the spin polarized state of the clusters is summarized as given below.
For any odd $N$ clusters, the stability of the spin-polarized state is higher than that of the spin-unpolarized state.
On the other hand, for most even $N$ clusters, the energy of the spin-unpolarized state is lower than that of the spin-polarized state.
However, several structures that have high denegeracy in the neighborhood of the HOMO in the electronic state are exceptions.
For a geometric structure with high symmetry, such as an icosahedron at $N=13$ and $N=55$,
cuboctahedron at $N=38$, capped icosahedron at $N=71$, and Marks decahedron at $N=75$, the energy levels around HOMO are highly degenerated.
Therefore, the high-spin states exhibit a higher stability than the low-spin states in systems such as an icosahedron
for Na$_{55}$ \cite{Itoh-2} and Cu$_{55}$, although the case is opposite for Ag$_{55}$.
However, for all of them, the energy differences between the high- and low-spin states are very small-0.033, 0.022, and 0.012 eV, respectively.
The system is stabilized by lowering the symmetry from $I_h$ to $C_i$ in the structures based on the Jahn-Teller theorem \cite{JT}.

\subsection{HOMO-LUMO gap, $E(N)$, and $v(N)/v(2)$ for the most and meta-stable clusters}

In section V. A., only the electronic structure of the most stable cluster structure is discussed.
In this section, the electronic structures of the most and meta-stable cluster structures are compared
to understand the relationship between the geometric and electronic structures of the clusters.
To save space, only the result of Na$_N$ is presented.
FIG. 10 shows the HOMO-LUMO gap ((a) and (b)), $\Delta _{2}E(N)$ (a'), $\delta E(N)$ ((c) and (d)), and $v(N)/v(2)$ ((e), and (f))
of the most and meta-stable structures of Na$_N$ clusters for $2\leq N \leq 22$ and $15 \leq N \leq 75$, respectively.
The values of the most stable structures are connected by a line.
Several important features are observed from the comparison of the most and meta-stable structures, as described below.

A close correlation is observed between the HOMO-LUMO gap and $\Delta _{2}E(N)$ of the most stable structure.
As shown in FIGS. 10 (a) and (a'), there is a strong correlation in the $N$ dependence between them for a range of $2\leq N\leq 21$.
Although the HOMO-LUMO gap considered in this study is the Kohn-Sham HOMO-LUMO gap,
originally, the HOMO-LUMO gap is a physical value that corresponds to the difference
between the ionization potential and the electron affinity in the system.
The ionization potential of a neutral charged cluster composed of $N$ atoms is defined as the energy difference
between the total energy of the neutral cluster energy-$E(N, n)$- and the one electron detached charged cluster-$E(N, n-1)$.
The electron affinity of a neutral charged cluster composed of $N$ atoms is defined as the energy difference
between the total energy of the neutral cluster-$E(N, n)$- and the one electron attached charged cluster-$E(N, n+1)$.
Therefore, the HOMO-LUMO gap of a cluster composed of $N$ atoms-$HLG(N)$ is defined as $HLG(N)\equiv E(N, n+1)+E(N, n-1)-2E(N, n)$.
On the other hand, the energy curvature of a cluster composed of $N$ atoms-$\Delta _{2}E(N)$-
is defined as $\Delta _{2}E(N)\equiv E(N+1, n)+E(N-1, n)-2E(N, n)$.
Although they are clearly different physical values, the forms of these two types of values are very similar.
The close correlation described above should be studied in detail in the future.

Generally, as shown in Figures 10 (a) and (b), the most stable structures exhibit a relatively large HOMO-LUMO gap.
Further, as shown in FIG. 9, a cluster that has a large HOMO-LUMO gap tends to have
a lower HOMO and higher LUMO in the electronic structures, as described in section V. A.
However, from a comparison with the energy differences between the most and meta-stable structures from $\delta E(N)$ shown in FIGS. 10 (c) and (d),
it is apparent that a large HOMO-LUMO gap is not a necessary condition for the most stable structure.
For example, in Na$_{4}$, the meta-stable structure III exhibits a larger HOMO-LUMO gap than that of the most stable structure I.
Although it is not shown in this paper, the same type of example can also be shown for Cu$_N$ and Ag$_N$.

Finally, the $N$ dependence of the normalized cluster volume $v(N)/v(2)$ of the most and meta-stable structures,
and the relation to the total energy and the HOMO-LUMO gap is noted.
As shown in FIGS. 10 (e) and (f), it is apparent that the cluster volume is directly dependent
on the structural type, namely, L, P, O, and C. From the system stability, as shown in FIGS. 10 (c) and (d),
it is apparent that the most stable structures do not necessarily have a minimum value of $v(N)/v(2)$.
Further, the $N$ dependence of $v(N)/v(2)$ for $N\geq 7$ does not change significantly
if energetically closed structural isomers to the most stable structures are also considered.
As shown in the figure, it is apparent that the relationship between the HOMO-LUMO gap
and $v(N)/v(2)$ for the most and meta-stable structures cannot be simplified. 
\section{Conclusions}
The $N$-dependent geometric structure, system stability, and electronic structures of Na$_N$, Cu$_N$, and Ag$_N$ metal clusters
are studied in detail for a range of $2\leq N\leq 75$ based on the density functional theory.
Strong similarities are observed between the three different metal clusters.
These similarities originate from the outermost is \textit{s} electron in the alkali and noble metal atoms composing the each cluster.
Much stronger similarities are observed between the two noble metal clusters for the benefit of \textit{d} electrons.
The most stable structures are the same for the three different metal clusters for approximately half the cluster sizes $N$ considered in this study.
Even if the most stable structures are different, the same type of structures are obtained if the meta-stable structures are also considered.
For all of the three clusters, the structural type of the most stable structure changes
in the order L $\rightarrow$ P $\rightarrow$ O $\rightarrow$ C with increasing $N$.
This structural type transition leads to a deviation from the monotonic increase in volume with $N$.
A remarkable similarity is also observed for the $N$ dependence of cluster energy $E(N)$ for the most stable geometric structures.
This similarity is related to the similarity in the electronic structures.
The amplitude of this energy difference is larger in the two noble metal clusters than in the alkali metal cluster.
This is attributed to the contribution of the $d$ electrons to the bonds.
The magic number is defined in the framework of total energy calclations for the first time.
In the case of Na$_N$, a semi-quantitative comparisson between the experimental abundance spectra \cite{Knight}
and the total energy calculations is carried out.
For the improvement of the agreement with the experimental result, several possiblities arise for the total energy calculations.
The changing aspects of the Kohn-Sham eigenvalues from $N=2$ to $N=75$ are presented for the three different metal clusters.
The features of the bulk density of states already appeared at $N=75$ for all of the three clusters.
With increasing values of $N$, the HOMO-LUMO gap clearly exhibits an odd-even alternation and converges to 0.
This alternation is a specific feature of alkali and noble metal clusters in which each element has one \textit{s} electron in the outermost shell.
It is also found that in some cases, the HOMO-LUMO gap is large at odd value of $N$ and small at even value of $N$.
Although there is similarity in the $N$ dependence of the HOMO-LUMO gap between the three metal clusters,
it is much stronger between the two noble metal clsuters.
The spatially localized \textit{d} electrons in the noble metals are energetically localized.
The growth aspect of the \textit{d} band below the Fermi level of the noble metal clusters with increasing $N$ is presented.
A good correspondence is obtained in the $d$ characteristic of the electronic states between the cluster composed of 75 atoms and the bulk metal.
%
\section*{Acknowledgements}
%
The authors gratefully acknowledge the kind hospitality at the Institute for Materials Research
and the staff of the Center for Computational Materials Science for allowing the use of the Hitachi SR8000/64 supercomputing facilities.
The authors are grateful to Prof. Bernd von Issendorff for sharing his results prior to publication.
M. I. deeply acknowledges valuable information about group theory from Dr Kenta Hongo
and molecular dynamics simulation for clusters from Dr Tamio Ikeshoji.
M. I. deeply acknowledges valuable discussions with Dr Hiroshi Yasuhara.
M. I. deeply acknowledges Dr Koichi Yoshizaki for allowing us to refer to his thesis prior to publication.
\\
\section*{Appendix}
%
\begin{table*} 
\begin{center}
\label{calc-expt}
\begin{tabular}{ccccccc}
\hline
\hline
\multicolumn{1}{c}{Element} & \multicolumn{1}{c}{Type of data} & \multicolumn{2}{c}{\textit{d} (\AA)} & \multicolumn{2}{c}{$E_{b}/N$ (eV/atom)} & \multicolumn{1}{c}{$B_{0}$ (10$^{11}$N/m$^2$)} \\ \cline{3-7}
\hline
\multicolumn{1}{c}{}        & \multicolumn{1}{c}{}             & Dimer & Bulk                         & Dimer & Bulk                             & \multicolumn{1}{c}{Bulk} \\
\hline
\multicolumn{1}{c}{Na} & This study & 3.07 & 3.64 & 0.38 & 1.07 & \multicolumn{1}{c}{0.074} \\ \cline{2-7}
\hline
\multicolumn{1}{c}{} & Expt. & 3.079 \cite{CRC} & 3.659 \cite{Kittel} & 0.379 \cite{CRC} & 1.113 \cite{Kittel} & \multicolumn{1}{c}{0.068 \cite{Kittel}} \\
\hline
\multicolumn{1}{c}{Cu} & This study & 2.22 & 2.57 & 1.14 & 3.53 & \multicolumn{1}{c}{1.396} \\ \cline{2-7}
\hline
\multicolumn{1}{c}{} & Expt. & 2.220 \cite{CRC} & 2.55 \cite{Kittel} & 0.915 \cite{CRC} & 3.49 \cite{Kittel} & \multicolumn{1}{c}{1.37 \cite{Kittel}} \\
\hline
\multicolumn{1}{c}{Ag} & This study & 2.58 & 2.95 & 0.89 & 2.53 & \multicolumn{1}{c}{0.876} \\ \cline{2-7}
\hline
\multicolumn{1}{c}{}  & Expt. & 2.531 \cite{expt} & 2.89 \cite{Kittel} & 0.831 \cite{CRC} & 2.95 \cite{Kittel} & \multicolumn{1}{c}{1.007 \cite{Kittel}} \\
\hline
\hline
\end{tabular}
\caption{The averaged nearest neighbor diatomic distance \textit{d}, binding energy per atom $E_{b}/N$,
and bulk modulus $B_{0}$ of dimers and bulk crystals for Na, Cu, and Ag in the equilibrium position evaluated
from the DFT \cite{HK, KS}-GGA(PW91 \cite{PW91}) calculations in this study and other studies based on several experiments are presented.}
\end{center}
\end{table*}
%

%
%
\begin{figure*}
\begin{center}
\includegraphics[width=15cm]{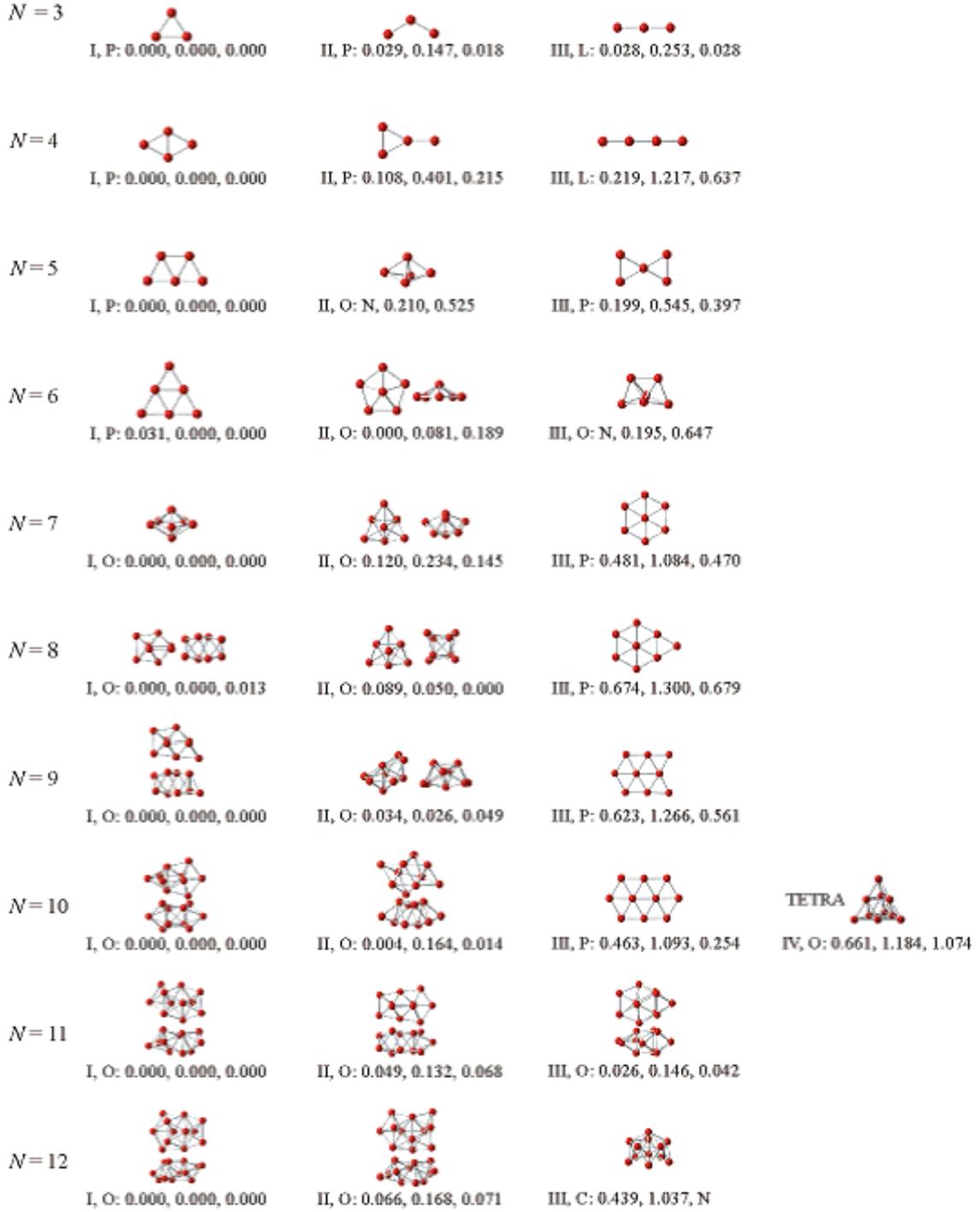}
\caption{\label{fig:structures-3-12}
The most and meta-stable structures
of Na$_N$, Cu$_N$, and Ag$_N$ clusters for $3 \leq N\leq 12$ are shown.
Almost all of the most and meta-stable structures of these clusters are similar.
Therefore, the cluster structures are represented by the structures of Cu$_N$ to save space in this figure.
The structures are classified according to their types using the notations I, II, III, and IV.
This classification obeys the stability order in the structural type of Cu$_N$.
The symbols L, P, O, and C used after the notations indicate linear, planar, opened, and closed structures, respectively.
Opened and closed structures are both three-dimensional structures.
An opened structure is defined as one without any atoms whose coordination number is greater than or equal to 11.
Other three-dimensional structures are difiend as closed structures.
The three values following the symbols L, P, O, and C represent
the relative total energies of the most stable structures of Na$_N$, Cu$_N$, and Ag$_N$, respectively.
A value of 0.000 is assigned to the most stable structure and the energy of meta-stable structures is expressed in units of electron volts.
The symbol N is used to denote a structure that cannot be identified.
The two different figures shown above show views of the same cluster from different angles.
The numerous highly symmetric structures, including Plato's polyhedron, are labeled as TETRA, OCTA, ICO, CUBO, and DECA,
and they respectively represents a tetrahedron, octahedron, icosahedron, cuboctahedron, and decahedron.
}
\end{center}
\end{figure*}
\begin{figure*}
\begin{center}
\includegraphics[width=15cm]{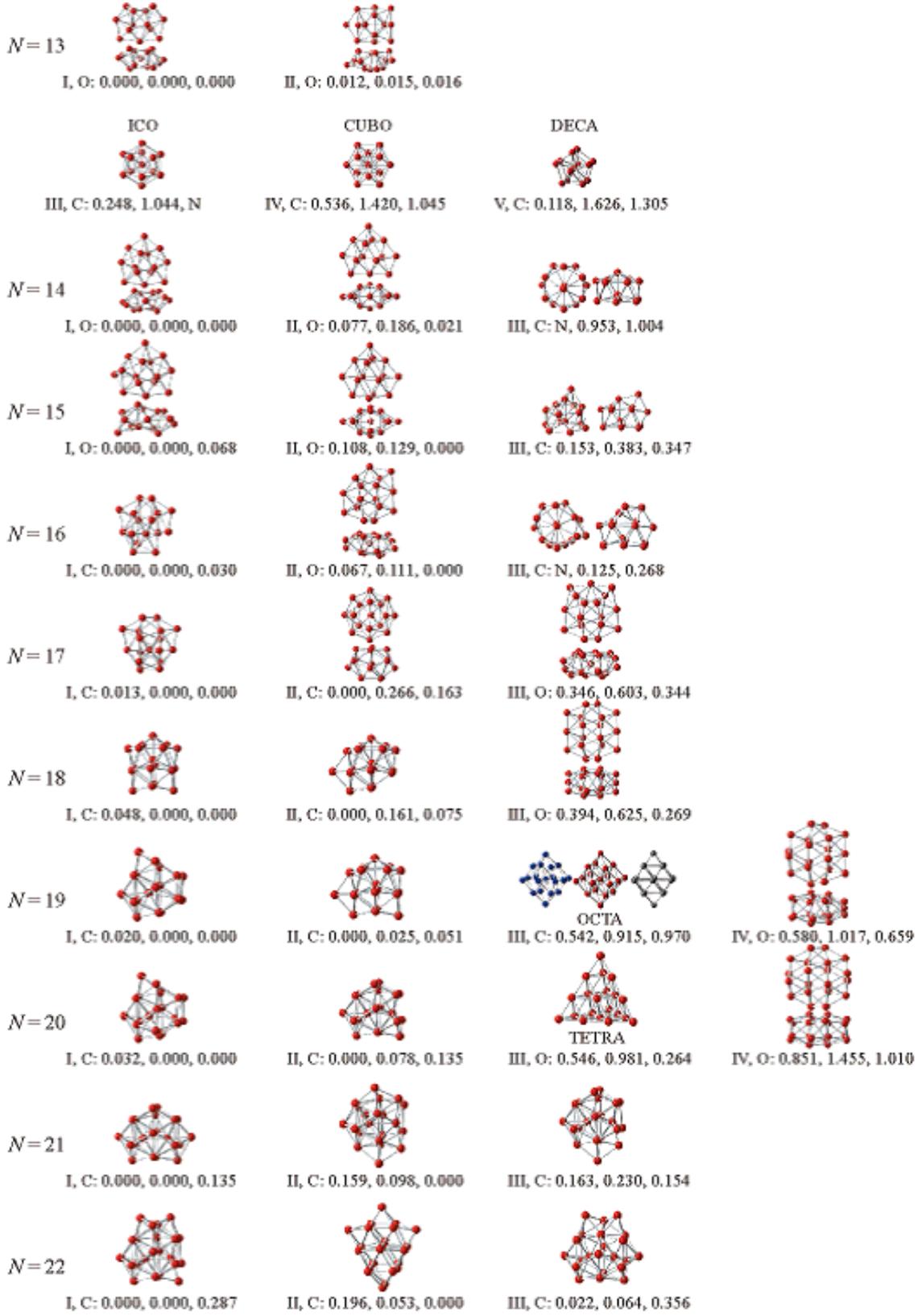}
\caption{\label{fig:structures-13-22}
The most and meta-stable structures
of Na$_N$, Cu$_N$, and Ag$_N$ clusters for $13 \leq N\leq 22$ are shown.
The notations used are the same as those in FIG.\ref{fig:structures-3-12}.
Some isomers of Na$_N$ and Ag$_N$ are also shown for cases where their structural deviations are relatively large from the relation of similarity.
The structures of Na$_N$ and Ag$_N$ are indicated in blue and silver, respectively.
%
}
\end{center}
\end{figure*}
\begin{figure*}
\begin{center}
\includegraphics[width=15cm]{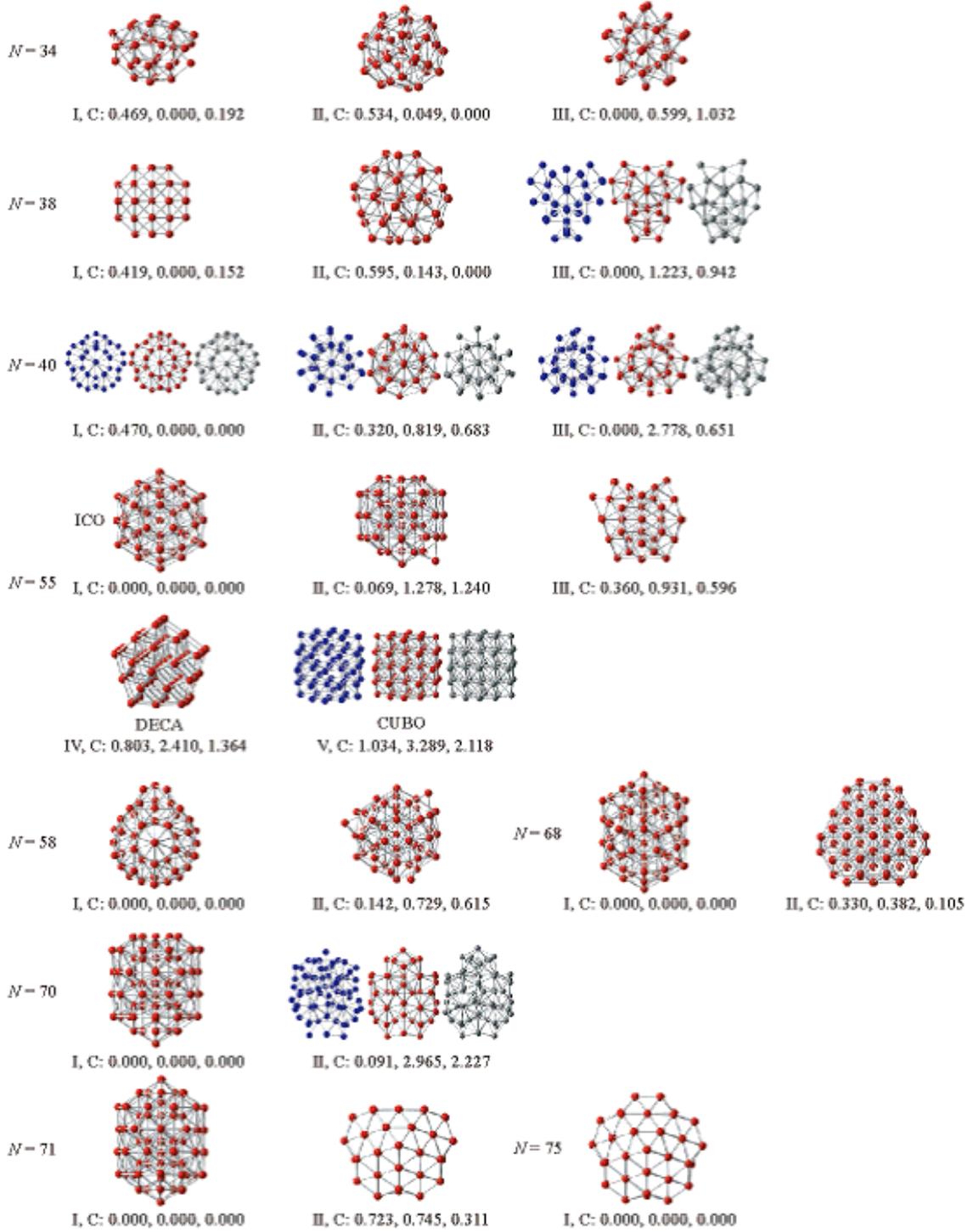}
\caption{\label{fig:structures-34-75}
The most and meta-stable structures of Na$_N$, Cu$_N$, and Ag$_N$ clusters for $34 \leq N\leq 75$ are shown.
The notations used are the same as those in FIGS.\ref{fig:structures-3-12} and \ref{fig:structures-13-22}.
}
\end{center}
\end{figure*}

\begin{figure*}
\begin{center}
\includegraphics[width=8cm]{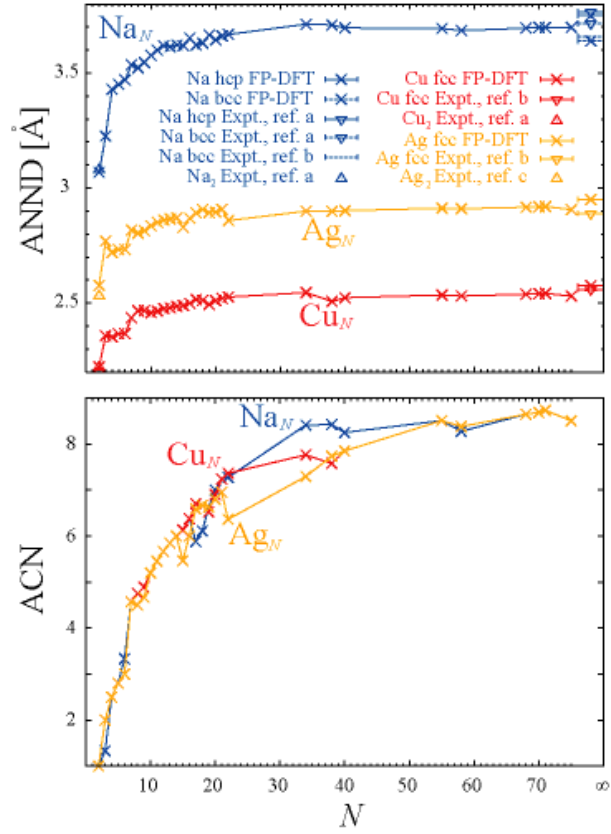}
\caption{\label{fig:ANND-ACN-PRB}
(a) shows the $N$ dependence of the averaged nearest neighbor distance (ANND) in units of angstroms,
and (b) shows the averaged coordination number (ACN) from atom to bulk
for Na$_N$, Cu$_N$, and Ag$_N$ for $1\leq N\leq 75$, and $\infty$ (bulk).
The ACN values of the bulk crystal - 12 (Na: hcp, Cu: fcc, Ag: fcc) and 8 (Na: bcc), are not shown in (b).
References a, b, and c, correspond to ref. \cite{CRC}, ref. \cite{Kittel}, and ref. \cite{expt}, respectively.
}
\end{center}
\end{figure*}
\begin{figure*}
\begin{center}
\includegraphics[width=12cm]{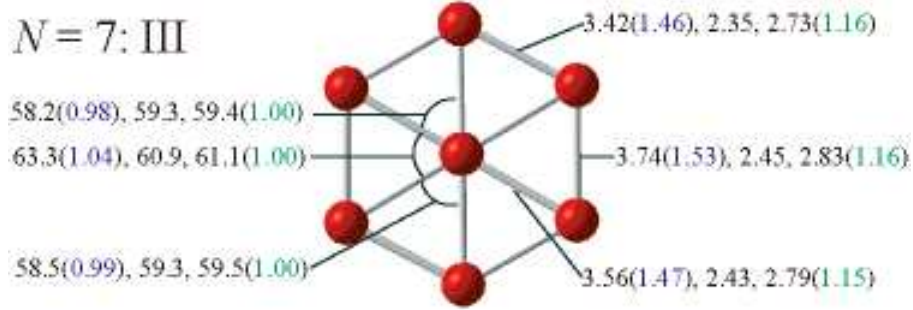}
\caption{\label{fig:structural-similarity-Na7-Cu7-Ag7}
The deviation from the relation of similarity shown in structure III of Na$_7$, Cu$_7$, and Ag$_7$ shown in Figure 1.
The values of the angles and interatomic distances are shown in the order of Na$_7$, Cu$_7$, and Ag$_7$.
Structure III of Cu$_7$ is shown as an example.
The angles and interatomic distances are expressed in units of degrees and angstroms, respectively.
The blue and green values in the parentheses represent the relative ratio of angle(distance)$_{Na_7}$/angle(distance)$_{Cu_7}$
and angle(distance)$_{Ag_7}$/angle(distance)$_{Cu_7}$, respectively.
}
\end{center}
\end{figure*}
\begin{figure*}
\begin{center}
\includegraphics[width=15cm]{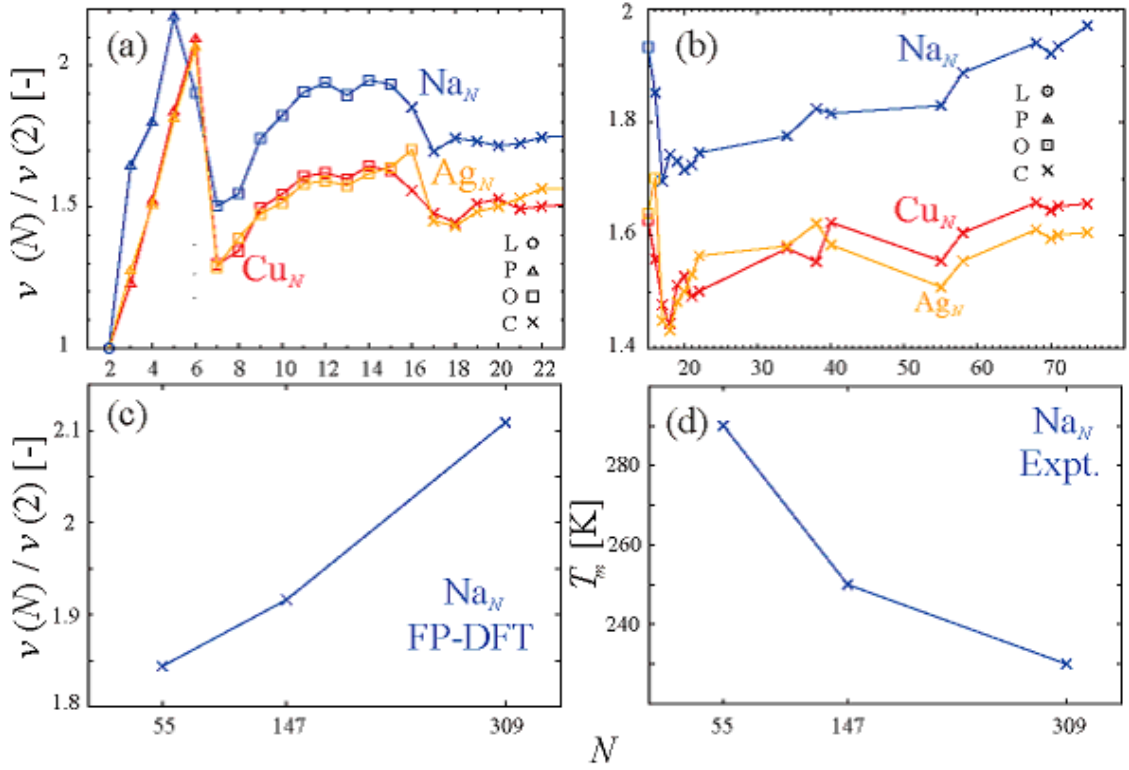}
\caption{\label{fig:VNV2-GS-PRB}
(a) and (b) show the $N$ dependence of the normalized cluster volume $v(N)/v(2)$ for the most stable structures
of Na$_N$, Cu$_N$, and Ag$_N$ for $2\leq N\leq 22$ and $15\leq N\leq 75$, respectively.
Each structure is denoted using the symbols L, P, O, and C.
(c) shows the $N$ dependence of $v(N)/v(2)$ for Na$_{N}$ ($N=55, 147$, and $309$),
while (d) shows the $N$ dependence of the melting point $T_m$ observed by Haberland \textit{et al.} \cite{Haberland-2005}
}
\end{center}
\end{figure*}
%
%
\begin{figure*}
\begin{center}
\includegraphics[width=12cm]{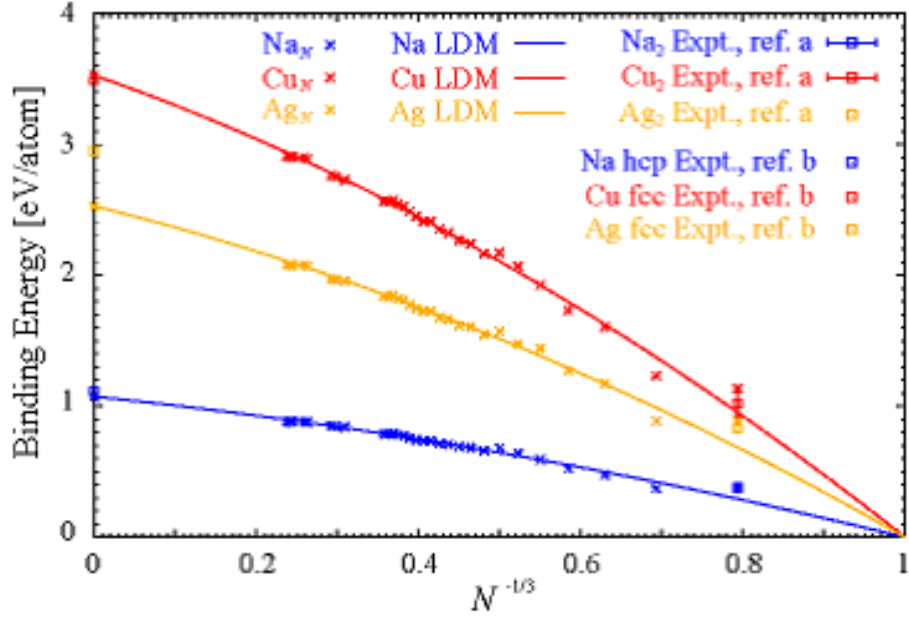}
\caption{\label{fig:BE-LDM}
The dependence of the binding energy of the ground state structures
of Na$_N$, Cu$_N$, and Ag$_N$ ($1 \leq N\leq 75, \infty$ (bulk)) on $N^{-\frac{1}{3}}$.
For each element, the liquid drop model (LDM) average is also shown.
References a and b correspond to ref. \cite{CRC} and ref. \cite{Kittel}, respectively.
}
\end{center}
\end{figure*}

\begin{figure*}
\begin{center}
\includegraphics[width=15cm]{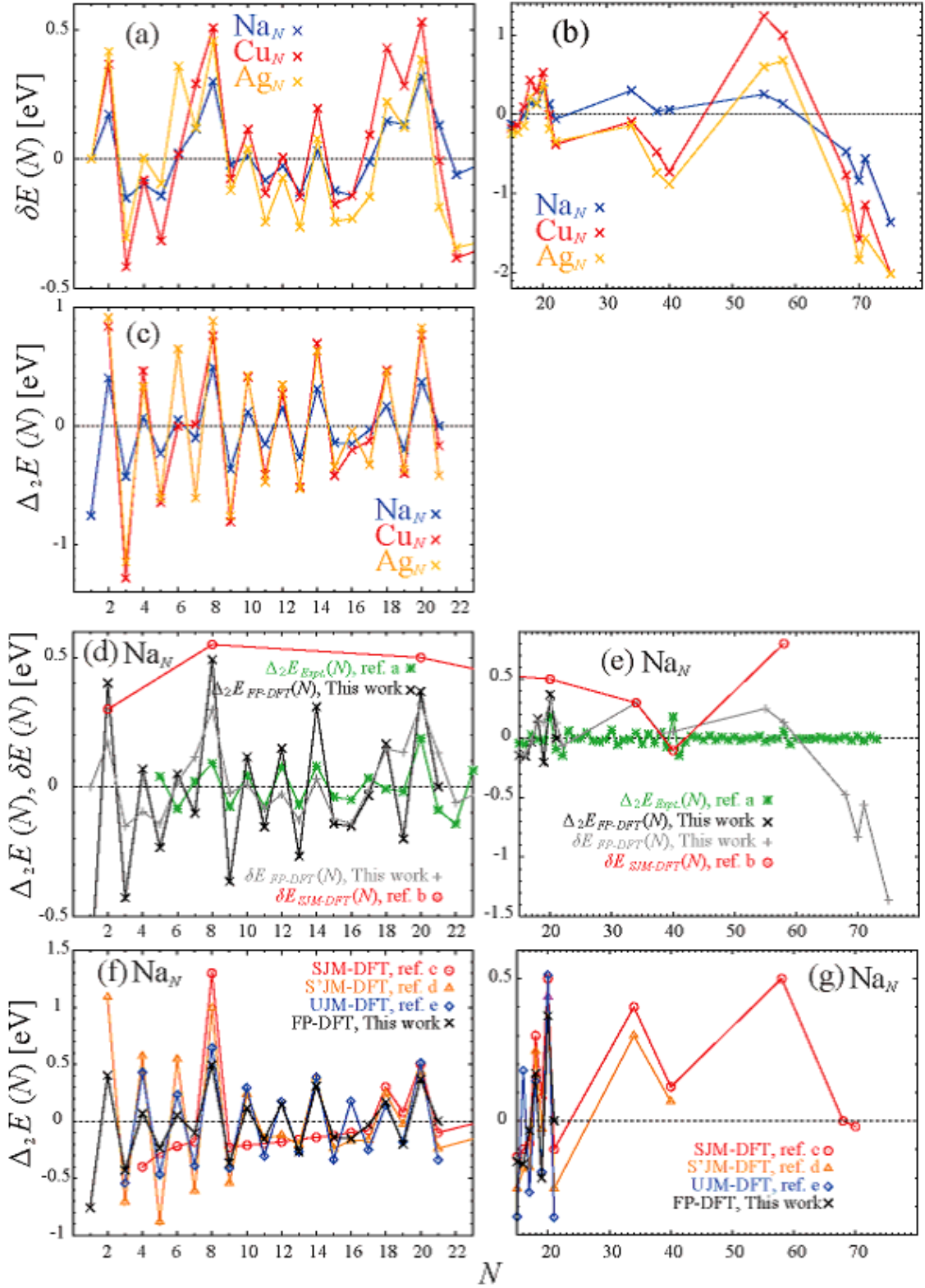}
\caption{\label{fig:deltaE-delta2E-PRB-submit}
(a) and (b) show the $N$ dependence of $\delta E(N)$ for the most stable structures
of Na$_N$, Cu$_N$, and Ag$_N$ clusters at $1 \leq N \leq 22$ and $15 \leq N \leq 75$, respectively.
(c) shows $\Delta _{2}E(N)$ for the most stable structures of Na$_N$, Cu$_N$, and Ag$_N$ at $1\leq N\leq 21$.
(d) and (e) shows the $N$ dependence of $\Delta _{2}E(N)=k_{B}T ln\frac{I(N)^{2}}{I(N+1)I(N-1)}$
of Na$_N$ as given by Knight \textit{et al.} (ref. a \cite{Knight}) at 700 kPa Ar in the cluster production step,
$\delta E(N)$ of the spherical jellium model (ref. b \cite{Genzken}) and first principles model,
and $\Delta _{2}E(N)$ of the first principles model for Na$_N$ clusters at $1 \leq N \leq 22$ and $15 \leq N \leq 75$, respectively.
For $\Delta _{2}E(N)$ value given by Knight \textit{et al.}, the value of the temperature is set to 800 K as measured
mearured in the nozzle channel before the cooling step in the experiment.
%
%
(f) and (g) show the $N$ dependence of $\Delta _{2} E(N)$ of Na$_N$ clusters
evaluated by various jellium models and the first principles model for $1 \leq N \leq 22$ and $15 \leq N \leq 75$, respectively.
References c, d, and e, correspond to ref. \cite{Chou}, ref. \cite{Ekardt-1988}, and ref. \cite{UJM}, respectively.
}
\end{center}
\end{figure*}
\begin{figure*}
\begin{center}
\includegraphics[width=15cm]{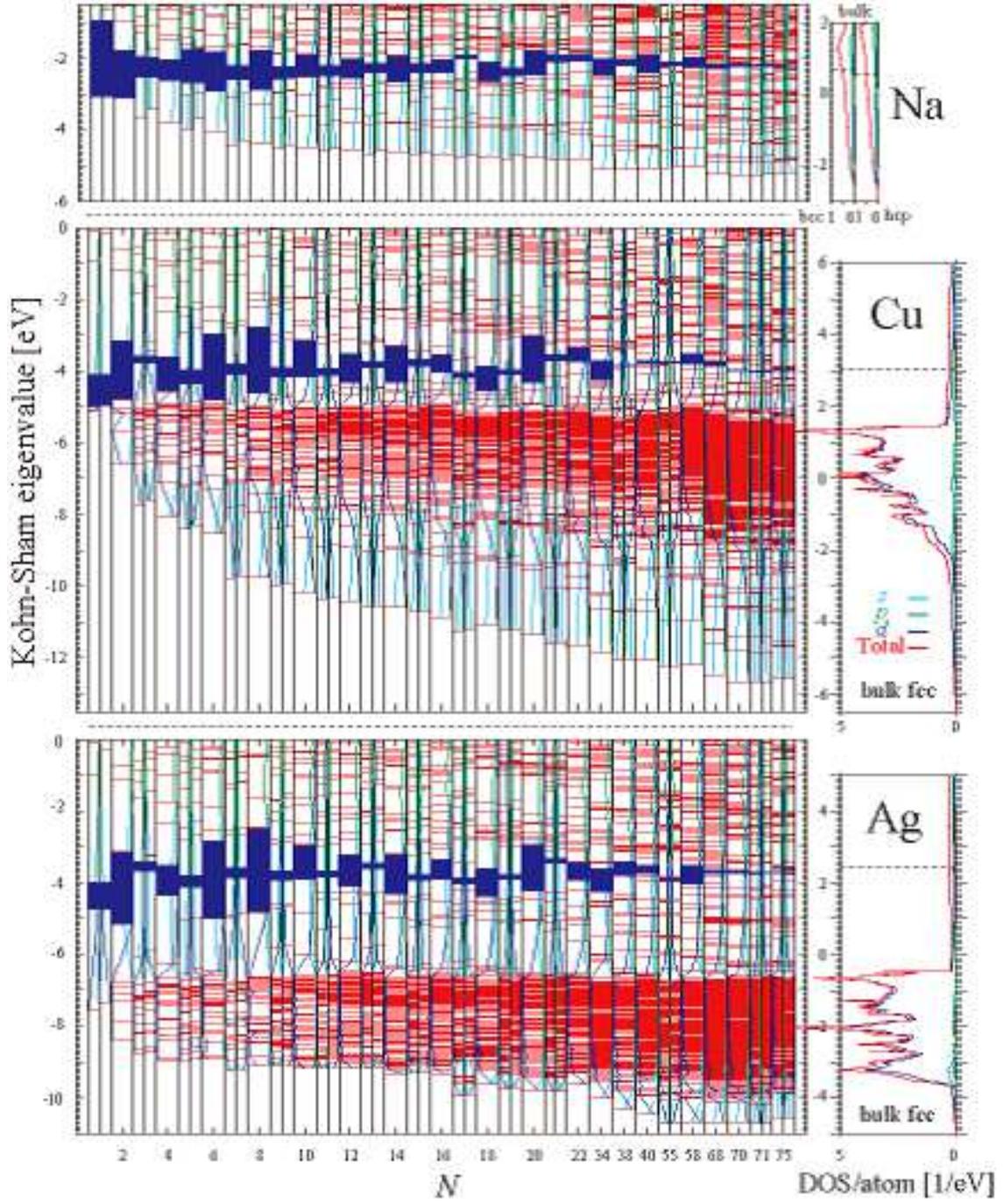}
\caption{\label{fig:Normarized-NaN-CuN-AgN-vs-KS-spd-bulk-PRB}
$N$ dependence of the Kohn-Sham eigenvalues from atom to bulk calculated
for the most stable structures of Na$_N$, Cu$_N$, and Ag$_N$ at $1\leq N\leq 75$ and $\infty$ (bulk). 
The red lines indicate occupied and unoccupied electronic energy levels.
For each spin-polarized system, the up spin states (left) and down spin states (right) are separated by a thin black line. 
The space between HOMO and LUMO is indicated in blue.
Each projected value of \textit{s}, \textit{p}, and \textit{d} to the Kohn-Sham state
is connected by water, green, and blue lines, respectively.
The density of states of each bulk crystal is shown to the right of each figure. Here, the Fermi level is represented by a dashed line.
The Fermi level of a bulk crystal is arranged near the HOMO of the clusters composed of 75 atoms.
Each colored line in the bulk crystal is the same as that in the case of clusters.
}
\end{center}
\end{figure*}
\begin{figure*}
\begin{center}
\includegraphics[width=15cm]{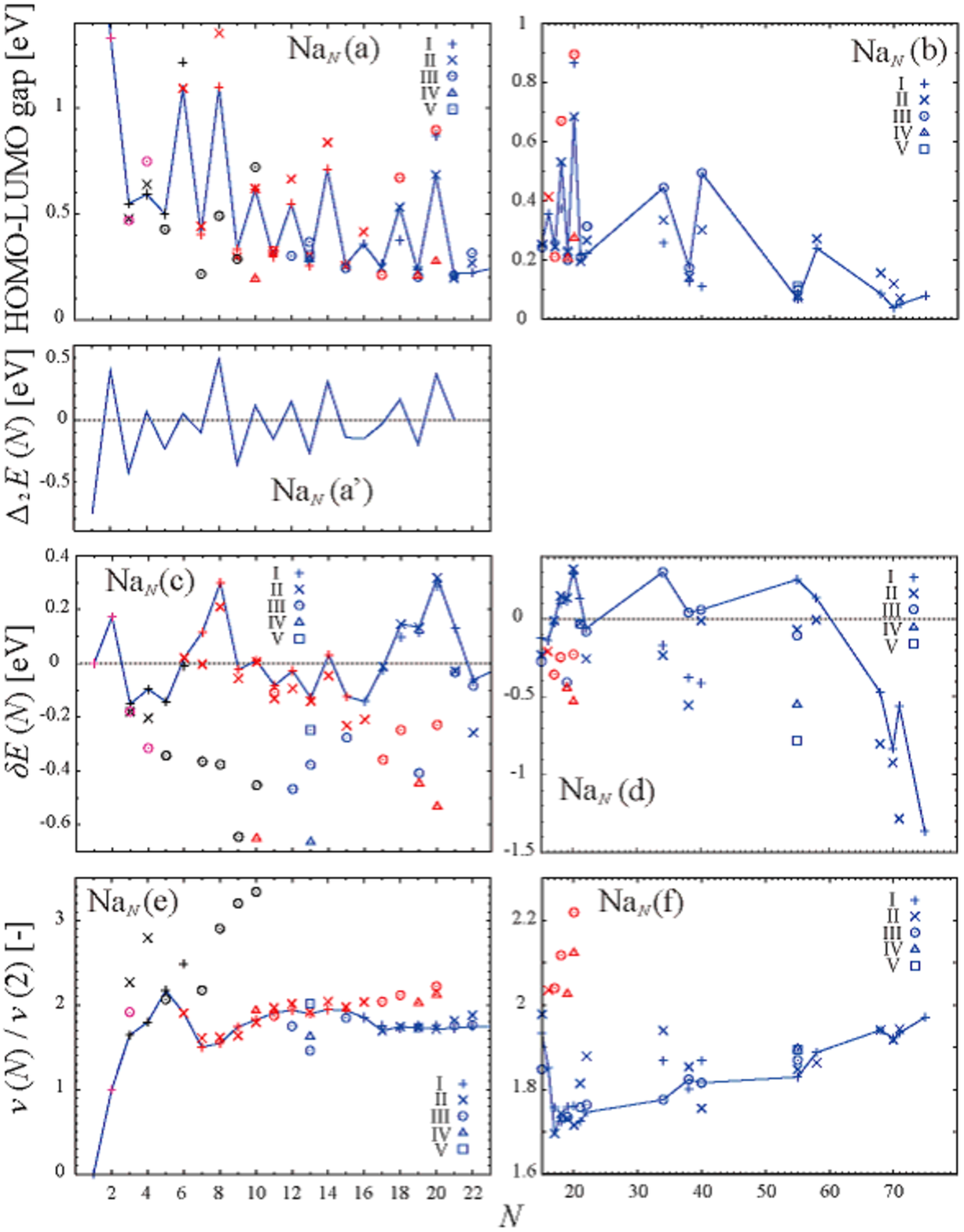}
\caption{\label{fig:Na-HLG-Delta2E-deltaE-VNV2-PRB-mod}
(a) and (b) show the $N$ dependence of the HOMO-LUMO energy gaps of Na$_N$
for the most and meta-stable structures at $1\leq N\leq 22$ and $15\leq N\leq 75$, respectively.
(a') shows the $N$ dependence of $\Delta _{2}E(N)$ for the most stable structures of Na$_N$ at $1\leq N\leq 21$.
(c) and (d) show the $N$ dependence of $\delta E(N)$ for Na$_N$
for the most and meta-stable structures at $1\leq N\leq 22$ and $15\leq N\leq 75$, respectively.
(e) and (f) show the $N$ dependence of $v(N)/v(2)$ for Na$_N$ for the most and meta-stable structures
at $1\leq N\leq 22$ and $15\leq N\leq 75$, respectively.
Each different type of structure, namely, linear (L), planar (P), opened (O), and closed (C),
is indicated in purple, black, red, and blue, respectively.
}
\end{center}
\end{figure*}

\begin{thebibliography}{99}
\bibitem{Knight-Ar}
From Figure 2 in reference \cite{Knight},
we confirmed that $\Delta _{2}E_{Expt.}(N)$ is not changed significantly by changing the Ar gas pressure in the first step in the experiment.
\end{thebibliography}
\end{document}